\documentclass[%
 aip,jcp,
 amsmath,amssymb,
preprint,%
]{revtex4-2}

\usepackage{graphicx}% Include figure files
\usepackage{dcolumn}% Align table columns on decimal point
\usepackage{bm}% bold math
\usepackage[T1]{fontenc}
\usepackage{mathptmx}

\usepackage{mathrsfs} 
\usepackage{physics}
\usepackage[caption=false]{subfig}
\usepackage{epstopdf}
\usepackage{multirow}
\usepackage{float}
\usepackage{xcolor}

\graphicspath{{./figures/}}

\begin{document}

\preprint{}

\title[]{Nucleus-electron correlation revising molecular bonding fingerprints 
from the exact wavefunction factorization
}
% Force line breaks with \\

\author{Ziyong Chen}
 %\altaffiliation{Department of Chemistry, The University of Hong Kong, Hong Kong SAR, China}%Lines break automatically or can be forced with \\
\author{Jun Yang}%
 \email{juny@hku.hk}
\affiliation{Department of Chemistry \\
The University of Hong Kong, Hong Kong SAR, P.R. China
%\\This line break forced with \textbackslash\textbackslash
}%

\date{\today}% It is always \today, today,
             %  but any date may be explicitly specified

\begin{abstract}
We present a novel theory and implementation for computing coupled electronic
and quantal nuclear subsystems on a single potential energy surface, moving
beyond the standard Born-Oppenheimer (BO) separation of nuclei and electrons. We
formulate an exact self-consistent nucleus-electron embedding potential from the
single product molecular wavefunction, and demonstrate that the fundamental
behavior of correlated nucleus-electron can be computed for mean-field electrons
that are responsive to a quantal anharmonic vibration of selected nuclei in a discrete
variable representation. Geometric gauge choices are discussed and necessary for
formulating energy invariant biorthogonal electronic equations.  Our method is
further applied  to characterize vibrationally averaged molecular bonding
properties of molecular energetics, bond length, protonic and electron density.
Moreover, post-Hartree-Fock electron correlation can be conveniently computed on
the basis of nucleus-electron coupled molecular orbitals, as demonstrated to
correlated models of second-order M{\o}llet-Plesset perturbation and full
configuration interaction theories. Our approach not only accurately quantifies
non-classical nucleus-electron couplings for revising molecular bonding properties, but
also provides an alternative time-independent approach for deploying non-BO
molecular quantum chemistry.
\end{abstract}

\maketitle

%\begin{quotation}
%The ``lead paragraph'' is encapsulated with the \LaTeX\ 
%\verb+quotation+ environment and is formatted as a single paragraph before the first section heading. 
%(The \verb+quotation+ environment reverts to its usual meaning after the first sectioning command.) 
%Note that numbered references are allowed in the lead paragraph.
%%
%The lead paragraph will only be found in an article being prepared for the journal \textit{Chaos}.
%\end{quotation}

\section{\label{sec:intro}Introduction}

The Born-Oppenheimer (BO) approximation\cite{BO1927} is the fundamental
cornerstone of modern electronic structure theories, providing computational
framework in which a broad range of chemical properties can be conveniently
computed. For instances, the BO approximation leads to the adiabatic electronic
potential energy surface (PES) with respect to nuclear positions, molecular
geometric structures obtained at PES stationary locations that are often
compared to experimentally resolved bond lengths and angles, the reaction
kinetics and pathways on a single PES where bonds are broken and made, infrared
spectroscopy from bond vibrations, and many others. However, since the BO
approximation accounts for only classical couplings for nucleus-electron pairs,
it breaks down when the non-classical nucleus-electron correlation contributes
significantly to such chemical processes involving light proton as
proton-coupled electron
transfer,\cite{PCET2001,PCET2008,PCET2010,PCET2012,PCET2015} nonadiabatic
quantum nuclear tunneling between close
PESs,\cite{ashfold2006role,roberts2012direct,xie2016nonadiabatic} and so on. The
presence of strongly correlated nucleus-electron motion may fundamentally alter
the density distributions of non-classical nuclei and electrons, giving rise to
different bonding characters and chemical connections averaged on nuclear
trajectories.

Many theoretical methods have been developed to address non-BO effects.
\cite{Yonehara2012,Habershon2013,Curchod2018} The most straightforward non-BO
scheme is the diagonal BO correction (DBOC) from the second derivative
coupling between adiabatic wavefunctions.\cite{DBOC2001,DBOC2003} The
multi-configurational time-dependent Hartree (MCTDH)\cite{Meyer1990,Meyer2000}
method determines the motion of quantal nuclei on several coupled PESs by
superposing the product states of electronic and nuclear wavefunctions.
The explicit correlated Gaussian (ECG) based non-BO theory,
uses the conventional Cartesian coordinates and ECG basis functions to build
many-body molecular wavefunction, which has been shown to converge
nucleus-electron correlation rapidly with the expansion
length.\cite{Kozlowski1993equivalent}
Quantum Monte Carlo (QMC) such as fixed-node diffusion QMC
(FN-DMC)\cite{tubman2014beyond,yang2015large} has been demonstrated to include a
nuclear wavefunction as the products of Gaussian functions on nuclear pairs.  
These methods have been demonstrated to be able to achieve high
accuracy, but are also challenged by large computational costs that limit their
application to non-BO effect of small systems.
\cite{NISHIZAWA2012142}

Moreover, significant efforts in developing non-BO formulism have been devoted
to the orbital representation of nuclei, originating from the idea of the
protonic wavefunctions using one-proton Slater-type functions centered on
heavier nuclei by Thomas.\cite{Thomas1969PR,Thomas1969CPL,Thomas1970} This
drove the introduction of more general Gaussian-type nuclear orbitals (NOs) on
an equal footing to electronic molecular orbitals (MOs), and led to the
development of various multicomponent-MO methods in which both electronic and
nuclear wavefunctions are computed simultaneously to incorporate
nucleus-electron correlation and nuclear quantum effects, as implemented at both
mean-field and correlated quantum chemical levels of theory including many-body
perturbation theory, coupled-cluster and configuration interaction
models.\cite{Tachikawa1998,tachikawa2000isotope,tachikawa2002multi,
nakai2003many,Bochevarov2004,ishimoto2009review}
The multicomponent MO provides a framework which is exact to non-BO problems if
it would be possible to perform the full configuration interaction (FCI) expansion
on the mean-field references for both electrons and
nuclei.\cite{Nakai2007nuclear}
Most recently, the multicomponent
nuclear-electronic orbital\cite{Hammes2020multicomponent} (NEO) method has
revitalized the idea of nuclear orbital to represent both classical and
quantized nuclei: a number of modern quantum chemistry variants from NEO
Hamiltonian and explicitly correlated wavefunction have been developed and
applied to non-BO
studies,\cite{Hammes2008,Hammes2017,pavosevic2020multicomponent} for methods
including density functional theory
(NEO-DFT),\cite{yu2020nuclear,yang2017development} constrained NEO-DFT 
(cNEO-DFT)\cite{Xu2020,Xu2020constrained,Xu2021molecular} and time-dependent DFT
(NEO-TDDFT),\cite{yang2018multicomponent,culpitt2019enhancing} coupled-cluster
singles and doubles (NEO-CCSD),\cite{Hammes2021multicomponent} orbital-optimized
second-order Møllet-Plesset (MP2) perturbation (NEO-OOMP2) and coupled-cluster
with doubles (NEO-OOCCD),\cite{pavosevic2020multicomponent} and the
complete-active space SCF (NEO-CASSCF).\cite{webb2002multiconfigurational}  As
the NEO Hamiltonian is nuclei-clamped for classical nuclei by assuming fixed
nuclear coordinates, the difficulties with rotational and translational degrees
of freedom are avoided.\cite{Hammes2020multicomponent} 
 Beyond multicomponent HF that neglects electron-proton correlation, 
Brorsen suggested that accurate protonic densities can be obtained form a truncated 
heat-bath CI expansion (HCI-NEO-CISDTQ)\cite{Brorsen2020quantifying} only if excitations up to quadruples are 
included, or the poor multicomponent HF orbitals are optimized in the presence of 
electron-proton correlation.\cite{Fajen2020separation,Fajen2021multi}

In principle, a rigorous separation of electronic and nuclear motion can be
represented in the Born-Huang (BH) expansion of the total molecular
wavefunction using complete adiabatic eigenstates of BO Hamiltonian. 
Alternative to BH, an exact probability decomposition was attempted by Hunter in
the early days\cite{Hunter1975}: the full molecular wavefunction
$\Psi(\mathbf{rs},\mathbf{R})$ was factorized into a single product of a non-BO
electronic wavefunction (as the conditional probability amplitude
$\Phi(\mathbf{rs},\mathbf{R})$ for electrons at $\bf{rs}$ in the presence of all
nuclei) and a nuclear wavefunction (as the marginal probability amplitude
$\chi(\mathbf{R})$ for nuclei at $\bf{R}$),
\begin{eqnarray}
 \Psi(\mathbf{rs},\mathbf{R}) = \Phi(\mathbf{rs},\mathbf{R})\chi(\mathbf{R}). \label{eq:CoupledProduct}
\end{eqnarray}
Recently, Gross et al. has shown that given partial normalization condition,
such an exact factorization exists and is unique for defining electronic and
nuclear subsystems, up to a phase factor. The exactly
factorized wavefunction has been investigated and validated for both
time-independent\cite{Cederbaum2013,Cederbaum2014,Gross2014,requist2016exact,li2018density}
and time-dependent\cite{Gross2010, Gross2012} non-BO simulations of several
model systems for which the original full problem is solved and
the factorization of Eq.~(\ref{eq:CoupledProduct}) is then inverted to obtain
electronic and nuclear subsystem
wavefunctions.\cite{Cederbaum2014vc,Gross2014n,Gross2017} A self-consistent
numerical approach for the time-dependent solution of coupled electron-ion
dynamics has been demonstrated to Shin-Metiu model system
recently.~\cite{Gossel2019}

Based on  Hunter's wavefunction probability interpretation for including non-BO
impacts, we propose a time-independent SCF method in which the
mean-field molecular wavefunction for quantum chemistry Hamiltonian is
factorized into a unique single product associated with non-BO electronic MOs
and the numerically exact vibrational wavefunction of nuclei. In this work, we
develop a computationally systematic and convenient approach of
self-consistently capturing non-classical nucleus-electron couplings for
uncorrelated electrons, and use resulting non-BO MOs and exact nucleus-electron
embedding potential to further build up electronic correlations in the well
established context of \textit{ab-initio} molecular quantum chemistry. By
computing and assessing non-BO effects on energies, chemical bond lengths,
electron and nuclear density distributions, we will demonstrate that our method
can accurately quantify nucleus-electron correlations which address
vibrationally averaged bonding patterns, from a mean-field single product
wavefunction.

\section{\label{sec:th}Theory}

We adopt the following notation in our formulation. The occupied, virtual and
general MOs are labeled by $\{i,j,k,\cdots\}$, $\{a,b,c,\cdots\}$ and
$\{p,q,r,\cdots \}$, respectively.  The MOs and their biorthogonal counterparts are
respectively denoted as $\psi_p$ and $\overline\psi_p$. The molecular geometry
is collectively signified by $\mathbf{R}$, and $\mathbf{r}$ and $\mathbf{s}$
represent the spatial and spin coordinates of electrons, respectively.
Generically, $\hat{O}$ and $\mathbf{O}$ are used to denote an operator and the
corresponding matrix form, respectively, with the latter composed of the matrix
elements by $O_{pq}$.

\subsection{\label{sec:factorization}Molecular wavefunction factorization}
Consider an H$^-$ anion that constitutes a minimum correlated non-BO atom.
Apparently, the three-body correlation emerges between the electron pair and the
proton, which complicates non-BO treatments within the hierarchy of correlated
quantum chemistry. Nonetheless, the original complex three-body problem can be
approximately cast into additive two-body problems in which the
electron-electron and electron-proton correlations can be separately
computed when a single PES dominates. To this end, we begin
with defining the full one-electron nucleus-electron wavefunction
$\widetilde\psi_i(\mathbf{r_js_j},\mathbf{R})$ in an exact factorization as follows,
\begin{equation} \widetilde{\psi}_i(\mathbf{r_js_j},\mathbf{R}) =
 \psi_i(\mathbf{r_js_j},\mathbf{R})\chi(\mathbf{R}) \label{eq:nonbomo} 
\end{equation} which is
analogous to that of the full many-electron molecular wavefunction of
Eq.~(\ref{eq:CoupledProduct}).
$\psi_i(\mathbf{r_js_j},\mathbf{R})$, which we term a non-BO electronic MO,
resembles the conditional probability amplitude of finding an electron in a
mean-field potential dressed in nucleus-electron correlation that explicitly
depends on nuclear coordinates.
The mean-field molecular wavefunction can be built by making a single product
Eq.~(\ref{eq:CoupledProduct})) between the non-BO electronic 
determinant $\Phi(\mathbf{rs},\mathbf{R})$ and
nuclear wavefunctions $\chi(\mathbf{R})$,
\begin{equation}
 \Phi(\mathbf{rs},\mathbf{R})
=\ket{\psi_1(\mathbf{r_1s_1},\mathbf{R}),\psi_2(\mathbf{r_2s_2},\mathbf{R}),\cdots,\psi_{N_e}(\mathbf{r_{N_e}s_{N_e}},\mathbf{R})},
\label{eq:Psi}
\end{equation}
for a free molecule of $N_e$ electrons.
This requires a normalization for $\Psi(\mathbf{rs},\mathbf{R})$ on both electronic and
nuclear coordinates,
\begin{eqnarray}
  \int\text{d}\mathbf{R}\braket{\Psi(\mathbf{rs},\mathbf{R})}{\Psi(\mathbf{rs},\mathbf{R})}_{\mathbf{rs}} = 1
\end{eqnarray}
and a partial normalization condition (PNC) for the determinant and thus non-BO
MOs at any molecular geometry $\mathbf{R}$ 
\begin{eqnarray}
 \braket{\psi_i(\mathbf{rs},\mathbf{R})}{\psi_i(\mathbf{rs},\mathbf{R})}_{\mathbf{rs}} = 1 \label{eq:PNC}
\end{eqnarray}
where $\ip{}{}_{\mathbf{rs}}$ indicates the normalization on the
electronic coordinates only.  This makes the unique decomposition of
Eqs.~(\ref{eq:nonbomo}) and~(\ref{eq:CoupledProduct}), up to only a gauge
transformation.  The coupled product of Eq.~(\ref{eq:CoupledProduct}) is to be
distinguished from the BO factorization 
$\Psi(\mathbf{rs},\mathbf{R})
\approx\Phi^\text{BO}(\mathbf{rs},\mathbf{R})\chi^\text{BO}(\mathbf{R})$ in
which $\Phi^\text{BO}(\mathbf{rs},\mathbf{R})$ must be a solution to the BO
electronic equation that parametrically depends on a geometry $\mathbf{R}$.
However, the BO product does not account for nucleus-electron correlation, which
can be recovered via the coupled product that encodes the mutual dependence
between electrons and nuclei. 
A tensor-product framework for constructing the
total wavefunction in molecular Hilbert space has been also demonstrated
recently.~\cite{sibaev2020molecular}

\subsection{\label{sec:potential}Exact nucleus-electron potential embedding
uncorrelated electrons}

In our approach, the full correlation between the nuclear and electronic
subsystems is determined via an SCF procedure in which the mutual response to
the electronic and nuclear impact from each other must be variationally
recovered by minimizing the total molecular energy functional $\mathscr{L}$ in
terms of $\Phi(\mathbf{rs},\mathbf{R})$ and $\chi(\mathbf{R})$, subject to the
normalization conditions in the electronic and nuclear coordinate space,
respectively.
\begin{widetext}
\begin{eqnarray}
\mathscr{L} &=& \int\text{d}\mathbf{R}\mel**{\Psi(\mathbf{rs},\mathbf{R})}{\hat{H}}{\Psi(\mathbf{rs},\mathbf{R})}_{\mathbf{rs}}
-\int\text{d}\mathbf{R}E_{el}(\mathbf{R})
\left[\braket{\Phi(\mathbf{rs},\mathbf{R})}{\Phi(\mathbf{rs},\mathbf{R})}_{\mathbf{rs}}-1\right]
\nonumber \\ 
&&- E\left[\braket{\chi(\mathbf{R})}{\chi(\mathbf{R})}_\mathbf{R}-1\right]
\label{eq:lagrang}
\end{eqnarray}
\end{widetext}
with the multipliers $E_{el}(\mathbf{R})$ and $E$ to enforce the PNC and
nuclear normalization conditions, respectively. The full molecular Hamiltonian
$\hat{H}$ is 
\begin{eqnarray}
  \hat{H} = \hat{H}_\mathrm{BO} + \hat{T}_n.
\end{eqnarray}
where $\hat{H}_\mathrm{BO}$ is the regular BO Hamiltonian.
$\hat{T}_n$ is the nuclear kinetic operator for all internal nuclear
motions, i.e., the atomic vibrations are separated from the continuum states
associated with the center of mass translation and the whole rotation of a free
molecule, with the latter providing only a constant energy shift.
%\begin{equation}
%  \hat{T}_n(\mathbf{R})=-\sum_{\nu=1}^{N_{n}} \frac{\nabla_{\nu}^{2}}{2 M_{\nu}}.
%\end{equation}
The spatial part of non-BO MO $\psi_i({\mathbf{r},\mathbf{R}})$ is expanded in
atomic orbitals (AO) $\phi_\alpha(\mathbf{r},\mathbf{R})$ through molecular
coefficients $C_{\alpha i}$ as an explicit function of nuclear positions.
\begin{eqnarray}
  \psi_i({\mathbf{r},\mathbf{R}}) = \sum_\alpha C_{\alpha i}(\mathbf{R})\phi_{\alpha}(\mathbf{r},\mathbf{R}).
\end{eqnarray}
For a single determinant wavefunction in Eq.~(\ref{eq:Psi}), the Lagrangian
energy of Eq.~(\ref{eq:lagrang}) can be cast into the form (see
Eq.~(\ref{eq:lagrangmo}) in Appendix~\ref{sec:derivation}) in terms of
biorthogonal non-BO MOs $\psi_i(\mathbf{r},\mathbf{R})$, and variationally
minimized with respect to $C_{\alpha i}(\mathbf{R})$, $\chi(\mathbf{R})$, the
multipliers $\tilde\epsilon_{ij}(\mathbf{R})$ and the total energy $E$, The derivation is
presented in Appendix~\ref{sec:derivation} and yields the non-BO coupled
electronic and nuclear equations in terms of a nucleus-electron correlation
potential for electrons and an electronic energy potential for nuclei as
follows, respectively,
\begin{eqnarray}
 %\left(\frac{}{}\mathbf{F} + \mathbf{V}^\mathrm{cp}(\chi)\right)\mathbf{c}_i
 %= \frac{\epsilon_i(\mathbf{R})}{|\chi|^2} \mathbf{S} \mathbf{c}_i
 \left(\frac{}{}\hat{F}_\mathrm{BO}+\hat{V}^\mathrm{cp}_\chi
+\frac{\chi^\dagger(\mathbf{R}) \hat{T}_n \chi(\mathbf{R})}{|\chi(\mathbf{R})|^2}
 \right)\ket{\psi_i}
&=& \frac{\epsilon_i(\mathbf{R})}{|\chi(\mathbf{R})|^2}\ket{\psi_i},
\label{eq:working1} \\
  %\left(-\sum_\nu\frac{1}{2M_\nu}\grad_\nu^2 + E_{el}(\mathbf{R}) \right)\chi = E\chi. 
\left(\hat{T}_n + E_{el}(\mathbf{R}) \right)\chi &=& E\chi.
\label{eq:working2}
\end{eqnarray}
$\hat{F}_{\mathrm{BO}}$ is the BO Fock operator.
The vibrational kinetic energy
$\chi^\dagger(\mathbf{R})\hat{T}_n(\mathbf{R})\chi(\mathbf{R})/|\chi(\mathbf{R})|^2$
provides constant shift to $\epsilon_i(\mathbf{R})$  for each geometry and can
be excluded from Eq.~(\ref{eq:working1}).
The value of the multiplier $E$ is the exact total molecular energy, and
the electronic $E_{el}(\mathbf{R}) =
\mel{\Phi(\mathbf{r},\mathbf{R})}{\hat{H}}{\Phi(\mathbf{r},\mathbf{R})}_{\mathbf{r}}
-\sum_\nu\frac{1}{M_\nu}\mel{\Phi(\mathbf{r},\mathbf{R})}{\grad_\nu}{\Phi(\mathbf{r},\mathbf{R})}_{\mathbf{r}}\cdot\grad_\nu$
defines the non-BO PES on which nuclei move, but also depends on nuclear
wavefunction. 
$\hat{V}^\mathrm{cp}_\chi$ is the nonlinear nucleus-electron embedding operator
that depends on the embedded one-electron states $\ket{\psi_i}$ and
$\bra{\overline \psi_j}$, their nuclear derivatives and the nuclear wavefunction
$\chi$, 
\begin{equation}
 \hat{V}^\mathrm{cp}_\chi = -\sum_\nu\frac{1}{M_\nu}\left(
\frac{\grad_\nu\chi}{\chi} \cdot\grad_\nu + \frac{\grad^2_\nu}{2}
+\sum_{j}\braket{\overline{\psi}_j}{\grad_\nu \psi_j}\cdot \grad_\nu
-\ket{\grad_\nu \psi_j} \cdot \bra{\overline{\psi}_j}\grad_\nu\right).
\label{eq:vc}
\end{equation}
 
Here, the first term $\frac{\grad_\nu\chi}{\chi} \cdot\grad_\nu$ is important to
capture the derivative coupling with nuclear motion, proportional to the
first geometric gradients of both electronic and nuclear wavefunctions.  The
remaining three terms account for
$\mel{\Phi(\mathbf{r},\mathbf{R})}{\hat{T}_n}{\Phi(\mathbf{r},\mathbf{R})}_{\mathbf{r}}$
which formally confirms the SCF DBOC contribution,\cite{Sellers1984adiabatic} but
in terms of $\Phi(\mathbf{r},\mathbf{R})$ that is self-consistently coupled to
the nuclear motion via $\hat{V}_\chi^\mathrm{cp}$. 
$\hat{V}^\mathrm{cp}_\chi$ represents the exact correlation potential with
nuclei over their internal coordinates $\nu$ (e.g., vibrational modes) for
embedding uncorrelated electrons non-classically, which must necessitate a
self-consistent procedure to solve Eqs.~(\ref{eq:working1}) for
$\psi_i(\mathbf{r},\mathbf{R})$ and (\ref{eq:working2}) for $\chi(\mathbf{R})$.
Our approach is therefore termed nucleus-electron coupled self-consistent field
(NECSCF) method which must converge both the PES $E_{el}(\mathbf{R})$ and the
total molecular energy $E$. 
Eqs.~(\ref{eq:working1}) and (\ref{eq:vc}) are solved for MOs to build a
Slater determinant, and the generalization to multiconfigurational electronic
wavefunction is feasible by adopting a linear combination of configurations in
Eq. (\ref{eq:CoupledProduct}).  The resulting MOs
$\psi_i(\mathbf{r},\mathbf{R})$ naturally lead to post-Hartree-Fock treatment of
correlated electrons in the presence of the exact quantum embedding potential
$\hat{V}^\mathrm{cp}(\chi)$ from nuclear subsystems that are governed on a
single PES. We thus further implemented the NECSCF-based second-order
M{\o}ller-Plesset perturbation (NECSCF-MP2) theory and the full configuration
interaction (NECSCF-FCI).

The nucleus-electron embedding potential $\hat{V}^\mathrm{cp}(\chi)$ must be
solved in the presence of the derivative operator
$\mathbf{\grad}_\nu=\frac{\partial~~}{\partial \mathbf{R}_v}$ in
Eq.~(\ref{eq:vc})
for including the non-BO relaxation of an electron that correlates with nuclei.
As $\hat{F}=\hat{F}_{BO}+\hat{V}_\chi^\mathrm{cp}$ is not self-adjoint
due to $\frac{\grad_\nu \chi}{\chi}\cdot\grad_\nu$,
the following biorthogonal equation is solved for non-Hermitian
system\cite{rosas2018bi} for its conjugate Hamiltonian,
\begin{equation}
%\hat{F}\ket{\psi_i}=\frac{\epsilon_i(\mathbf{R})}{|\chi|^2}\ket{\psi_i}
\hat{F}^\dagger\ket{\overline{\psi}_i}=\frac{\overline{\epsilon}_i(\mathbf{R})}{|\chi|^2}\ket{\overline{\psi}_i}.
\end{equation}
with the biorthogonality
$\braket{\overline{\psi}_i}{\psi_j}=\braket{\psi_i}{\overline{\psi}_j}=\delta_{ij}$
and $\overline{\epsilon}^\ast_i(\mathbf{R})=\epsilon_i(\mathbf{R})$.

\subsection{\label{sec:gauge}Geometric gauge choice making energy invariant}
For general biorthogonal non-BO MOs that lose complex conjugation, the
first-derivative $\braket{\overline{\psi}_i}{\grad_\nu\psi_j}$ may be no
longer anti-symmetric 
%$\braket{\overline{\psi}_i}{\grad_\nu\psi_j}\neq\braket{\grad_\nu\overline{\psi}_j}{\psi_i}$ 
which may break the invariance of electronic energy upon an unitary rotation
within the respective internal space on $\ket{\psi_i}$ and
$\ket{\overline{\psi}_i}$. 
The elements of the geometric derivative vectors $\mathbf{A}^{(\nu)}$ and
$\mathbf{\overline A}^{(\nu)}$ are 
\begin{eqnarray}\label{eq:anti}
 A_{pq}^{(\nu)}&=&\braket{\overline{\psi}_p}{\grad_\nu{\psi}_q} \\
 \overline A_{pq}^{(\nu)}&=&\braket{\psi_p}{\grad_\nu\overline{\psi}_q},
\end{eqnarray}
Based on the derivative biorthogonality from
\begin{equation}\label{eq:biorth}
\grad_\nu\braket{\overline{\psi}_p}{\psi_q}=0,
\end{equation} 
$\mathbf{A}$ and $\mathbf{\overline A}$ must be related to each other as,
\begin{eqnarray}
 \mathbf{\overline A}^{\dagger(\nu)} &=& -\mathbf{A}^{(\nu)}.
\end{eqnarray}
Many geometric gauges that fulfill the biorthogonality condition
of Eq.~(\ref{eq:biorth}) can be envisioned. In the present work, 
we  construct  a set  of legitimate $U_{ij}^{(\nu)}$ and
$\overline{U}_{ij}^{(\nu)}$ that are composed of symmetric and anti-symmetric
components among the occupied MOs,
\begin{eqnarray}
 U_{ij}^{(\nu)} &=&
-\frac{1}{2}\left(
\mathbf{\overline{C}}_i^\dagger\mathbf{S}^{0\nu}\mathbf{C}_j
+\mathbf{\overline{C}}_j^\dagger\mathbf{S}^{0\nu}\mathbf{C}_i\right) 
-\frac{1}{2}\left(
\mathbf{\overline C}_i^\dagger\mathbf{S}^{\nu 0}\mathbf{C}_j
-\mathbf{\overline{C}}_j^\dagger\mathbf{S}^{\nu 0}\mathbf{C}_i\right), 
\label{eq:gaug21}\\
 \overline{U}_{ij}^{(\nu)} &=&
-\frac{1}{2}\left(
\mathbf{C}_i^\dagger\mathbf{S}^{0\nu}\mathbf{\overline C}_j
+\mathbf{{C}}_j^\dagger\mathbf{S}^{0\nu}\mathbf{\overline C}_i\right) 
-\frac{1}{2}\left(
\mathbf{C}_i^\dagger\mathbf{S}^{\nu 0}\mathbf{\overline C}_j
-\mathbf{C}_j^\dagger\mathbf{S}^{\nu 0}\mathbf{\overline C}_i\right).
\label{eq:gaug22}
\end{eqnarray}
Above, the derivative overlaps are
$S_{\alpha\beta}^{0\nu}=\braket{\phi_\alpha}{\grad_\nu\phi_\beta}$ and
$S_{\alpha\beta}^{\nu 0}=\braket{\grad_\nu\phi_\alpha}{\phi_\beta}$.
Similar constructions can be drawn to virtual MOs.
This gives rise to the anti-symmetric $A_{ij}^{(\nu)}$ and
$\overline{A}_{ij}^{(\nu)}$,
\begin{eqnarray}
A_{ij}^{(\nu)} &= & 
-\frac{1}{2}\left(
\mathbf{\overline{C}}_j^\dagger\mathbf{S}^{0\nu}\mathbf{C}_i
-\mathbf{\overline{C}}_i^\dagger\mathbf{S}^{0\nu}\mathbf{C}_j\right) 
-\frac{1}{2}\left(
\mathbf{\overline C}_i^\dagger\mathbf{S}^{\nu 0}\mathbf{C}_j
-\mathbf{\overline{C}}_j^\dagger\mathbf{S}^{\nu 0}\mathbf{C}_i \right),
\label{eq:A21} \\
\overline A_{ij}^{(\nu)} &= & 
-\frac{1}{2}\left(
\mathbf{{C}}_j^\dagger\mathbf{S}^{0\nu}\mathbf{\overline C}_i 
-\mathbf{C}_i^\dagger\mathbf{S}^{0\nu}\mathbf{\overline C}_j\right)
-\frac{1}{2}\left(
\mathbf{C}_i^\dagger\mathbf{S}^{\nu 0}\mathbf{\overline C}_j
-\mathbf{C}_j^\dagger\mathbf{S}^{\nu 0}\mathbf{\overline C}_i\right) 
\label{eq:A22} 
\end{eqnarray}
Obviously the diagonal elements must be zero
\begin{equation} 
A_{ii}^{(\nu)} = \overline{A}_{ii}^{(\nu)} = 0.
\end{equation}
As we will further show in Appendix~\ref{sec:unitary}, any unitary rotation
among all occupied MOs adds merely a phase factor to the total molecular
wavefunction $\Psi(\mathbf{rs},\mathbf{R})$, leaving both electronic and
molecular energies invariant. This is essential to orbital localization and
diabatization techniques based on NECSCF MOs as derived from the electronic
equation.

Moreover, another gauge transformation is possible by combining the Hermitian
and anti-Hermitian terms as follows
\begin{eqnarray}
 U_{ij}^{(\nu)} &=&
-\frac{1}{2}\left(
\mathbf{\overline{C}}_i^{\dagger}\mathbf{S}^{0\nu}\mathbf{C}_j
+\mathbf{C}_i^\dagger\mathbf{S}^{\nu 0}\mathbf{\overline{C}}_j \right) 
-\frac{1}{2}\left(
\mathbf{\overline{C}}_i^{\dagger}\mathbf{S}^{\nu 0}\mathbf{C}_j 
-\mathbf{{C}}_i^\dagger\mathbf{S}^{0\nu}\mathbf{\overline{C}}_j\right),
\label{eq:gaug31}\\
 \overline{U}_{ij}^{(\nu)} &=&
-\frac{1}{2}\left(
\mathbf{\overline{C}}_i^\dagger\mathbf{S}^{\nu 0}\mathbf{C}_j
+\mathbf{C}_i^{\dagger}\mathbf{S}^{0\nu}\mathbf{\overline{C}}_j \right)
-\frac{1}{2}\left(
\mathbf{C}_i^{\dagger}\mathbf{S}^{\nu 0}\mathbf{\overline C}_j 
-\mathbf{{\overline C}}_i^\dagger\mathbf{S}^{0\nu}\mathbf{ C}_j \right),
\label{eq:gaug32}
\end{eqnarray}
giving the following anti-Hermitian derivative vectors ,
\begin{widetext}
\begin{eqnarray}\label{eq:A31} 
A_{ij}^{(\nu)} &=\overline A_{ij}^{(\nu)} &= 
-\frac{1}{2}\left(
\mathbf{\overline{C}}_i^{\dagger}\mathbf{S}^{\nu 0}\mathbf{C}_j 
-\mathbf{{C}}_i^\dagger\mathbf{S}^{0\nu}\mathbf{\overline{C}}_j
\right) 
-\frac{1}{2}\left(
\mathbf{C}_i^{\dagger}\mathbf{S}^{\nu 0}\mathbf{\overline C}_j 
-\mathbf{{\overline C}}_i^\dagger\mathbf{S}^{0\nu}\mathbf{ C}_j 
\right).
\end{eqnarray}
\end{widetext}
A stronger condition can be therefore examed by enforcing all derivative
elements $A_{ij}^{(\nu)}=0$,
\begin{eqnarray}
\mathbf{\overline C}^\dagger_i\left(\mathbf{S}^{\nu 0}-\mathbf{S}^{0\nu}\right)\mathbf{C}_j
=-\mathbf{C}^\dagger_i\left( \mathbf{S}^{\nu 0}-\mathbf{S}^{0\nu}\right)\mathbf{\overline{C}}_j.
\label{eq:x1}
\end{eqnarray}
Apparently, the condition in Eq.~(\ref{eq:x1}) is Hermitian and leads to energy
invariant formulation, but may be difficult to occur simultaneously for general
many-electron molecules.  Detailed numerical studies on its possibility to the
localization/diabatization of NECSCF MOs will be exploited in our future work. 

\subsection{\label{sec:working}Working equation and approximation}

Under the transformations in Eqs.~(\ref{eq:gaug21}) and~(\ref{eq:gaug22}),
the embedding operator $\hat{V}^\mathrm{cp}_\chi$ can be reduced to
\begin{widetext}
\begin{equation}
 \hat{V}^\mathrm{cp}_\chi = -\sum_\nu\frac{1}{M_\nu}\left(
\frac{\grad_\nu\chi}{\chi} \cdot\grad_\nu + \frac{\grad^2_\nu}{2}
-\sum_{j}\ket{\grad_\nu \psi_j} \cdot
\bra{\overline{\psi}_j}\grad_\nu\right)
\label{eq:vcwork}
\end{equation}
\end{widetext}
with the last two terms accounting for the contribution from nuclear kinetic
operator. As shown in Appendix~\ref{sec:derivation}, $\hat{V}^\mathrm{cp}_\chi$ is
form-invariant upon an unitary rotation among occupied MOs.
By employing the identity operator
$\mathbf{I}=\sum_p \ket{\psi_p}\bra{\overline{\psi}_p}$ in the full
spectrum of non-BO MOs, 
\begin{eqnarray}
 \grad^2_\nu \ket{\psi_i} &=& \grad_\nu\cdot\left(\sum_p
\ket{\psi_p}\ip{\overline{\psi}_p}{\grad_\nu \psi_i}\right) 
\nonumber \\
&=& \sum_p\ket{\grad_\nu\psi_p}\cdot \ip{\overline{\psi}_p}{\grad_\nu \psi_i}
+\sum_p\ket{\psi_p}\grad_\nu\cdot \ip{\overline{\psi}_p}{\grad_\nu \psi_i}
\end{eqnarray}
there is
\begin{widetext}
\begin{eqnarray}\label{eq:kinetic}
&& \frac{\grad^2_\nu}{2} -\sum_{j}\ket{\grad_\nu \psi_j}
\cdot\bra{\overline{\psi}_j}\grad_\nu 
 \nonumber \\
=&& \grad^2_\nu -\sum_{j}\ket{\grad_\nu \psi_j}
\cdot\bra{\overline{\psi}_j}\grad_\nu - \frac{\grad^2_\nu}{2}
\nonumber \\
=&&\left( \sum_p \ket{\grad_\nu \psi_p}\cdot \bra{\overline\psi_p}\grad_\nu
+ \sum_p \ket{\psi_p}\grad_\nu\cdot\bra{\overline\psi_p}\grad_\nu \right)
-\sum_{j}\ket{\grad_\nu \psi_j} \cdot\bra{\overline{\psi}_j}\grad_\nu 
- \frac{\grad^2_\nu}{2}
\nonumber \\
=& & 
\sum_a \ket{\grad_\nu \psi_a}\cdot \bra{\overline\psi_a}\grad_\nu
+ \sum_p \ket{\psi_p}\grad_\nu\cdot\bra{\overline\psi_p}\grad_\nu 
- \frac{\grad^2_\nu}{2}.
\end{eqnarray}
\end{widetext}
The integration of Eq.~(\ref{eq:kinetic}) between $\overline\psi_i$ and $\psi_j$ over
their electronic coordinates yields the following element,
which is termed $Y^{(\nu)}_{ij}$,
\begin{eqnarray}
%\sum_a \braket{\overline\psi_i}{\grad_\nu \psi_a}\cdot\braket{\overline\psi_a}{\grad_\nu\psi_j}
%+ \grad_\nu\cdot\braket{\overline\psi_i}{\grad_\nu\psi_j}.
Y^{(\nu)}_{ij}&=&\sum_a A_{ia}^{(\nu)}\cdot A_{aj}^{(\nu)} + \grad_\nu \cdot A_{ij}^{(\nu)}
- \braket{\overline\psi_i}{\frac{\grad^2_\nu}{2}\psi_j} 
\nonumber \\
&=& \sum_a A_{ia}^{(\nu)}\cdot A_{aj}^{(\nu)} + \grad_\nu \cdot A_{ij}^{(\nu)}
-\frac{1}{2}\left( \grad_\nu\cdot \braket{\overline\psi_i}{\grad_\nu\psi_j} 
-\braket{\grad_\nu\overline\psi_i}{\grad_\nu\psi_j}
\right)
\nonumber \\
&=& \sum_a A_{ia}^{(\nu)}\cdot A_{aj}^{(\nu)} + \frac{1}{2}\grad_\nu \cdot A_{ij}^{(\nu)}
+\frac{1}{2}\braket{\grad_\nu\overline\psi_i}{\grad_\nu\psi_j}
\end{eqnarray}
Here the first term is quite small and  thus neglected in our solver of
electronic equation, as the product  $A_{ia}A_{aj}$ is inversely proportional to the product
of the difference between occupied and virtual orbital energies
$\frac{1}{(\epsilon_a -\epsilon_i)(\epsilon_j -\epsilon_a)}$.
A linearization of the second term is  made by using the first-order Taylor
expansion of the derivative $A_{ij}^{(\nu)}$ with respect to the
vibrational coordinates,
\begin{widetext}
\begin{equation}\label{eq:approx1}
\grad_\nu\cdot A_{ij}^{(\nu)} \approx \grad_\nu\cdot
\left[ A_{ij}^{(\nu)}\right]_{\mathbf{R}_0} +
\grad_\nu \cdot \sum_{\nu'} {\nu'} \left[\grad_{\nu'} \cdot A_{ij}^{(\nu)}\right]_{\mathbf{R}_0}
= \left[\grad_{\nu} \cdot A_{ij}^{(\nu)}\right]_{\mathbf{R}_0}
\end{equation}
\end{widetext}
which indicates that the divergence of the NECSCF first-derivative vector can be
approximated by that at the unperturbed geometry where the BO approximation is
assumed. 
Since $\grad_\nu \cdot A_{ij}^{(\nu)}$ makes no direct contribution to
the electronic energy due to the diagonal  $A_{ii}^{(\nu)}=0$ and
the off-diagonal contribution from the BO-based
$\left[A_{ij}^{(\nu)}\right]_{\mathbf{R}_0}$ to the embedding operator is real,
and usually very small compared to $\frac{\grad_\nu\chi}{\chi}\cdot A_{ij}^{(\nu)}$,
the divergence term $\grad_\nu\cdot A_{ij}^{(\nu)}$ is not important and thus
neglected as well in solving NECSCF electronic states. The last contribution
$\braket{\grad_\nu\overline\psi_i}{\grad_\nu\psi_j}$ is also usually neglected.
Finally, only
$\frac{\grad_\nu\chi}{\chi}\cdot\grad_\nu$ operator is practically important and
implemented in the AO basis for solving NECSCF one-electron states.  The
$Y_{ij}^{(\nu)}$ contributions that are not included for SCF solution can be
simply patched to NECSCF orbital energies using the converged NECSCF MOs.  
 
\begin{eqnarray}
\mathbf{V}_\chi^\mathrm{cp} &=& -\sum_{\nu} \frac{1}{M_{\nu}}
\frac{\grad_\nu\chi(\mathbf{R})}{\chi(\mathbf{R})}
\left[\mathbf{S}(\grad_\nu \mathbf{C}) \mathbf{\overline C}^{\dagger}
\mathbf{S}+\mathbf{S}^{0\nu}\right],
\label{eq:vmat1}
\\
\mathbf{V}_\chi^{\dagger\mathrm{cp}} &=& \sum_{\nu} \frac{1}{M_{\nu}}
\frac{\grad_\nu\chi^\dagger(\mathbf{R})}{\chi^\dagger(\mathbf{R})}
\left[\mathbf{S}(\grad_\nu \mathbf{\overline C}) \mathbf{C}^{\dagger}
\mathbf{S}+\mathbf{S}^{0\nu}\right]
\label{eq:vmat2}
\end{eqnarray}

The coupled perturbed Hartree-Fock (CPHF) is formulated for biorthogonal NECSCF
system to compute $\grad_\nu \mathbf{C}$ and $\grad_\nu \mathbf{\overline C}$,
based on the first-order geometric relaxation $\mathbf{U}^{(\nu)}$ and
$\mathbf{\overline{U}}^{(\nu)}$ of one-electron state embedded in the
non-Hermitian nucleus-electron potential.
$\mathbf{U}^{(\nu)}$ and $\mathbf{\overline{U}}^{(\nu)}$ are related through the
derivative biorthogonality between the eigenstates of $\hat{F}$ and
$\hat{F}^\dagger$, 
\begin{eqnarray} 
\mathbf{U}^{(\nu)}+\mathbf{\overline{U}}^{\dagger(\nu)}+\mathscr{S}^{(\nu)} &=& 0, 
\label{eq:Ubiorthonormal1}
%\mathbf{U}^{\dagger(\nu)}+\mathbf{\overline U}^{(\nu)}+\mathscr{\overline S}^{(\nu)} &=& 0,
%\label{eq:Ubiorthonormal2}
\end{eqnarray}
where in the MO basis
\begin{eqnarray}
\mathscr{S}^{(\nu)}&=&\mathbf{\overline C}^\dagger\left(\mathbf{S}^{0\nu}+\mathbf{S}^{\nu 0}\right)\mathbf{C}.
%\mathscr{\overline S}^{(\nu)}&=&\mathbf{C}^\dagger\left(\mathbf{S}^{0\nu}+\mathbf{S}^{\nu 0}\right)\mathbf{\overline C}.
\end{eqnarray}
Apparently, as $\mathbf{\overline{U}}^{\dagger(\nu)}$ is not assumed to be
conjugated with $\mathbf{U}^{(\nu)}$ which normally holds in BO methods, both
the occupied-virtual relaxation ($U^{(\nu)}_{ia}$) and virtual-occupied
relaxation ($U^{(\nu)}_{ai}$) must be explicitly solved from the CPHF equations,
\begin{widetext}
\begin{eqnarray}\label{eq:CPHFE}
\left[\epsilon_{q} - \epsilon_{p}\right]U_{pq}^{(\nu)} =&& 
\sum_k^\text{occ}\expval{pk\Vert qk}^{(\nu)}  +
\mathscr{H}^{(\nu)}_{pq} - \mathscr{S}^{(\nu)}_{pq}\epsilon_{q} 
-\sum_{jk}^{\mathrm{occ}}\mathscr{S}_{kj}^{(\nu)}\expval{pj\Vert
qk} \nonumber \\
&& + \sum_k^{\text{occ}}\sum_b^\text{vir}\left[U_{bk}^{(\nu)}\expval{pk\Vert qb} 
- U^{(\nu)}_{kb}\expval{pb\Vert qk} - \mathscr{S}_{kb}^{(\nu)}\expval{pb\Vert qk}
\right]
\end{eqnarray} 
\end{widetext}
where the one-electron derivative core Hamiltonian is
$\mathscr{H}^{(\nu)}_{pq} = 
\mathbf{\overline C}_p^{\dagger}\mathbf{G}_\mathrm{NECSCF}^{(\nu)}\mathbf{C}_q$.
Once $U_{ia}^{(\nu)}$ and $U_{ai}^{(\nu)}$ are solved, the relaxations
$\overline{U}_{ia}^{(\nu)}$ and $\overline{U}_{ai}^{(\nu)}$ are thus computed
using Eq.~(\ref{eq:Ubiorthonormal1}).

It should be pointed out that the derivative core Hamiltonian
$\mathbf{G}_\mathrm{NECSCF}^{(\nu)}$ contains the NECSCF contribution from the
nucleus-electron embedding potential. According to the
first-order response of $\frac{\grad_\nu
\chi(\mathbf{R})}{\chi(\mathbf{R})}\cdot \grad_\nu$ operator in AO matrix form,
\begin{widetext}
\begin{eqnarray}
\grad_{\nu}\sum_{\nu'} \frac{1}{M_{\nu'}}\frac{\grad_{\nu'}\chi(\mathbf{R})}{\chi(\mathbf{R})}\cdot \mathbf{A}^{(\nu')} 
& = &
\sum_{\nu'}\frac{1}{M_{\nu'}}\grad_\nu\left[\frac{\grad_{\nu'}\chi(\mathbf{R})}{\chi(\mathbf{R})}\right]\cdot \mathbf{A}^{(\nu')}
+ \sum_{\nu'}\frac{1}{M_{\nu'}}\frac{\grad_{\nu'}\chi(\mathbf{R})}{\chi(\mathbf{R})}\cdot \grad_\nu \mathbf{A}^{(\nu')}.
\end{eqnarray}
\end{widetext}
Using similar analysis according to
Eq.~(\ref{eq:approx1}), $\grad_\nu \mathbf{A}^{(\nu')}\approx \Big[\grad_\nu
\mathbf{A}^{(\nu')}\Big]_\mathbf{R_0}$ is unimportant and neglected to avoid second geometric
derivative. 
Further assuming the vibrational product state for independent vibrational wavefunctions
$\chi_\nu(\mathbf{R})$, there is
\begin{equation}
\frac{\grad_\nu\chi(\mathbf{R})}{\chi(\mathbf{R})} =
\frac{\grad_\nu\chi_\nu(\mathbf{R})}{\chi_\nu(\mathbf{R})},
\end{equation}
and therefore
\begin{equation}
\grad_\nu\left[\frac{\grad_{\nu'}\chi(\mathbf{R})}{\chi(\mathbf{R})} \right]
=
\delta_{\nu\nu'}\grad_\nu\left[\frac{\grad_{\nu'}\chi_{\nu'}(\mathbf{R})}{\chi_{\nu'}(\mathbf{R})}\right].
\end{equation}
We then admit the following working form of
$\mathbf{G}_\mathrm{NECSCF}^{(\nu)}$,
\begin{equation}\label{eq:gnecscf}
 \mathbf{G}_\mathrm{NECSCF}^{(\nu)}=\mathbf{G}_\mathrm{BO}^{(\nu)}
- \sum_\nu \frac{1}{M_\nu}\grad_\nu
\left[ \frac{\grad_\nu\chi_\nu(\mathbf{R})}{\chi_\nu(\mathbf{R})} \right]
\cdot \mathbf{A}^{(\nu)}
\end{equation}
with $\mathbf{G}_\mathrm{BO}^{(\nu)}$ the derivative of the regular core
Hamiltonian accounting for the electronic kinetic and classical nucleus-electron attraction
energies computed in the BO framework. Analytical first- and second-derivatives
$\grad_\nu \chi_\nu$ and $\grad_\nu^2 \chi_\nu$ of vibrational wavefunction
$\chi_\nu$ are implemented (Appendix~\ref{sec:fghgrad}).

Two simple scenarios can be analyzed based on
Eqs.~(\ref{eq:vmat1}),~(\ref{eq:vmat2}) and~(\ref{eq:gnecscf}) to exam the
significance of correlated nucleus-electron behaviour.
Around equilibrium structure, the ground vibrational wavefunction $\chi_\nu$ can
be approximated as Gaussian function for which
$\frac{\grad_\nu\chi_\nu(\mathbf{R})}{\chi_\nu(\mathbf{R})}$ is proportional to
vibrational displacement and $\grad_\nu\left[\frac{\grad_\nu\chi_\nu(\mathbf{R})}{\chi_\nu(\mathbf{R})} \right]$ 
proportional to frequency $\omega_\nu$ of this mode, i.e., higher-frequency
modes and moderately stretched bonds tend to strengthen nucleus-electron
correlation.  However, at the dissociation region where the electronic potential
becomes flat, the ground vibrational wavefunction carries the landscape
asymptotically approaching an exponential function for which the NECSCF
contribution is eventually zero due to
$\grad_\nu\left[\frac{\grad_\nu\chi_\nu(\mathbf{R})}{\chi_\nu(\mathbf{R})}
\right]\approx 0$, i.e., the nucleus-electron correlation becomes much weakened.
\begin{figure}[h]
     \includegraphics[width=\textwidth]{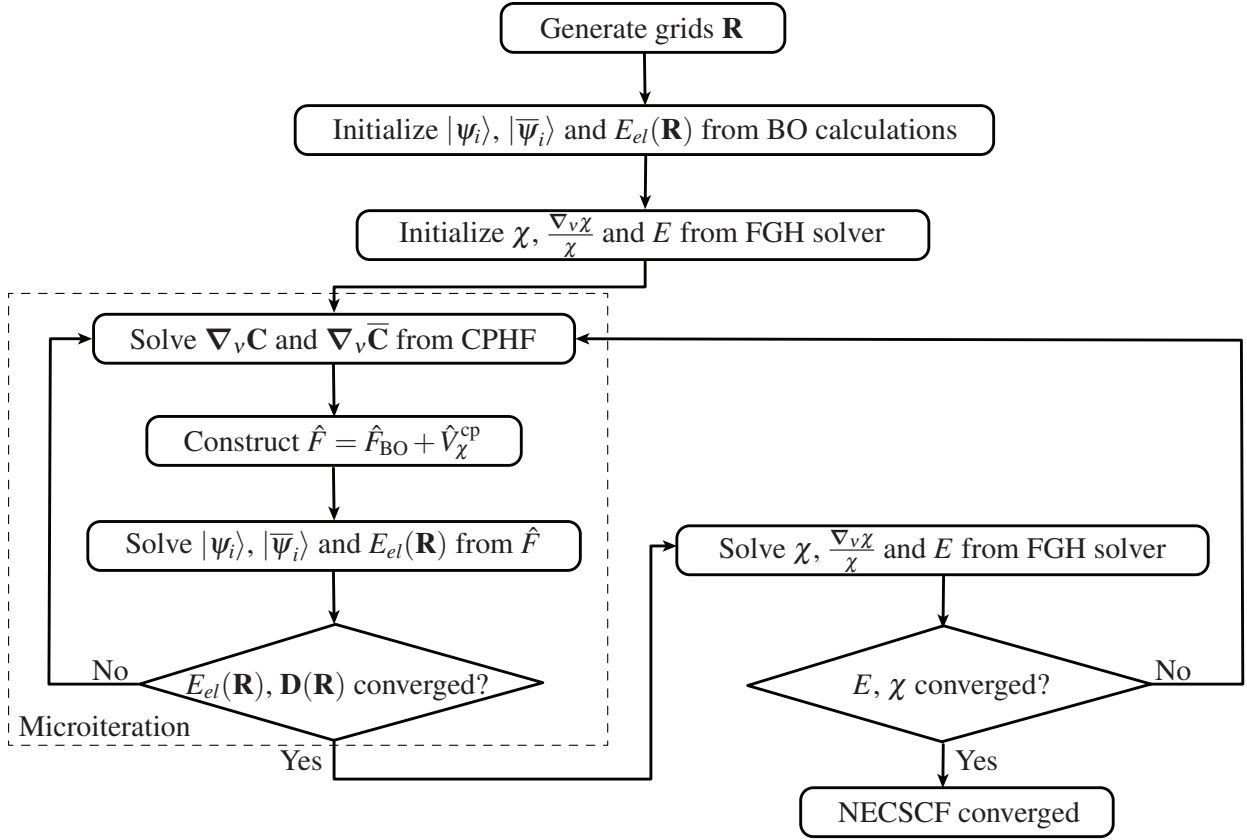}
     \caption{The SCF iteration in NECSCF implementation. The
electronic microiteration computes and converges $E_{el}(\mathbf{R})$,  the
electronic density matrix $\mathbf{D(R)}$ and the embedding potential in the
presence of $\frac{\grad_\nu\chi}{\chi}$ resulting from  the nuclear equation
that is solved in the previous macroiteration. Each microiteration solves both
NECSCF-based Hartree-Fock and coupled perturbed Hartree-Fock equations
iteratively, and macroiteration computes the molecular energy by exact
diagonalization. This two-step optimization proceeds until the numerical
convergence of both electronic and molecular energies within specified
criteria.}
     \label{fig:flow}
\end{figure} 

The nuclear wavefunction in Eq.~(\ref{eq:working2}) is numerically represented
in a discrete variable representation (DVR), e.g., on a set of uniform grid
points on
which Fourier Grid Hamiltonian\cite{FGH1989,FGH2003} (FGH) is fully
diagonalized.  When the nuclear wavefunction $\chi(\mathbf{R})$ becomes
oscillatory, for instance, for excited vibrations, the nodes at which
$\chi(\mathbf{R}_\mathrm{nodes})=0$ pose strong singularities in the embedding
potential $\mathbf{V}^\mathrm{cp}(\chi)$ that creates derivative discontinuities
(i.e., cusps) in $\psi_i(\mathbf{r},\mathbf{R}_\mathrm{nodes})$ and
singularities in $E_{el}(\mathbf{R}_\mathrm{nodes})$ through
Eq.~(\ref{eq:working1}).  To circumvent this problem, we solve
Eqs.~(\ref{eq:working1}) and (\ref{eq:working2}) piecewisely according to the
node positions of $\chi$;  on the other hand, the proximal region of
$\chi(\mathbf{R}_\mathrm{nodes})=0$ is where nucleus-electron correlations are
deemed strong.\cite{Czub1978, Gossel2019}

The SCF procedure of NECSCF implementation is depicted in
Fig.~\ref{fig:flow}.  The initial electronic orbitals are taken from standard BO
Hartree-Fock MOs.  An initial nuclear wavefunction  is constructed on the
uniform grid points of selected nuclear modes and computed using FGH solver for the
initial BO PES.  We adopt a two-step optimization in which microiteration cycles
for electronic wavefunction are carried out in the presence of
$\frac{\grad_\nu\chi}{\chi}$ solved from the previous NECSCF macroiteration.
The NECSCF convergence can be usually achieved in a reasonable number of
iterations for small molecules reported in this work (see
Fig.~\ref{fig:NECSCF_cycle} for numerical demonstration). 

\subsection{\label{sec:relation}Relation to variational search of exactly factorized wavefunction}

Having introduced the NECSCF method, one fundamental and interesting question
would be to ask whether, in principle, such an exact solution exists for a system
of coupled electrons and vibrations.  It can be proved that the minimization
search for optimal Slater determinant for the self-consistently coupled
electronic and nuclear equations does exist, however, on three stationary
conditions of the molecular energy with respect to variations of electron
density, nuclear wavefunction and paramagnetic current density
($\mathbf{j}(\mathbf{r},\mathbf{R})$).\cite{requist2016exact} While the NECSCF
equations can be solved for the conditional electronic wavefunction and marginal
nuclear wavefunction variationally, the last condition requires the direct
minimization of the energy with respect to paramagnetic current density to
uniquely determine the geometric vector potential
$\mathbf{A}_\mathrm{oo}^{(\nu)}$ and to account for induced electromagnetic
interactions with nuclei. Our NECSCF procedure leaves a freedom to choose
$\mathbf{A}_\mathrm{oo}^{(\nu)}$ from solving electronic and nuclear equations,
and we have discussed (Section~\ref{sec:gauge}) to fix this freedom by making
the gauge choice of $A_{ii}^{(\nu)}=0$ via a geometric relaxation of the
occupied NECSCF MOs. As a proxy to the third stationary condition, this choice
equally vanishes the paramagnetic current density as
$\mathbf{j}(\mathbf{r},\mathbf{R})\sim
\mathrm{Im}\tr(\mathbf{A}_\mathrm{oo}^{(\nu)})$ for Slater determinant
wavefunction. 

The impact of $\mathbf{j}(\mathbf{r},\mathbf{R})$ (or vector potential
$\mathbf{A}_\mathrm{oo}^{(\nu)}$) can be understood from the effective coupled
equations (Eqs.~(\ref{eq:effect_elec}) and (\ref{eq:effect_nuc}) in
Appendix~\ref{sec:effective}). The effective nuclear kinetic energy operator
$\hat{\widetilde T}_n$ (Eq.~(\ref{eq:tn2})) defines the kinetic energy of
electronically coupled nuclei that appear to move on an effective electronic
potential surface $\widetilde E_{el}(\mathbf{R})$ governing the effective
nuclear equation~Eq,~\ref{eq:effect_nuc}, as opposed to the original nuclear
equation (Eq.~(\ref{eq:working_nuc})).

\section{\label{sec:comp}Computational Details and Efficiency}

All electronic structure computations presented in this work were performed
using cc-pVTZ basis set for non-H atoms and aug-cc-pVTZ for H except for
basis set convergence tests.  Our NECSCF program is interfaced with the PySCF program
package\cite{PySCF} for accessing to one- and two-electron
integrals and their derivatives. The nuclear wavefunction was constructed on the
grid points that are
evenly spaced 0.005 {\AA} apart and sufficiently wide such that the tail of
resulting nuclear amplitudes decays below $10^{-5}$ au. 
For instance, 321 and 241 grid points were used for the ground vibrational level of 
H$_2^+$ and H$_2$, respectively. For molecules that contain heavy atoms, 
the number of grid points required is smaller than that for H$_2^+$ and H$_2$. 
To reduce wall-clock time, the parallel computing was enabled by allocating electronic SCF 
calculation on each nuclear grid to individual process within the microiteration.

\begin{figure}[h!]
\includegraphics[width=0.5\textwidth]{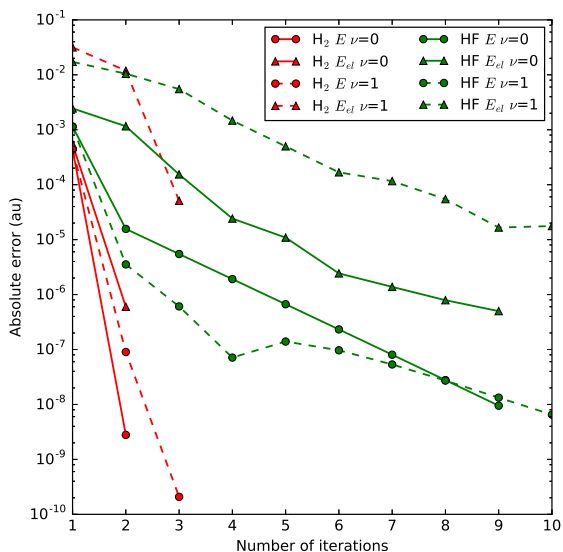}
 \caption{Comparisons of the NECSCF energy convergence performance
(macroiteration) of the total molecular energy ($E$) and non-BO electronic PES
($E_{el}(\mathbf{R})$) for H$_2$ and HF at  $\nu=0, 1$ vibrational levels.
The NECSCF is converged when the change of $E$ is within $10^{-8}$ 
au in the macroiteration.
The convergence criterion for the electronic
microiteration is met if the minimal change of the electronic energy
is below $10^{-4}$ au or the maximal number of micro-iterations exceeds 20.
The absolute energy change (au) shows energy updates of $\Delta E$ (filled
circle) and the maximal $\Delta E_{el}(\mathbf{R})$ (filled triangle)
amongst all grid points between two successive macroiteration cycles.}
\label{fig:NECSCF_cycle}
\end{figure} 

The iterative absolute energy updates are shown in Fig.~\ref{fig:NECSCF_cycle}
between two successive macroiterations for H$_2$ and HF molecules.  For such a
very simple H$_2$ molecule, the molecular energy $E$ converges in 2
macroiterative cycles for vibrational ground and that for $\nu=1$ level requires
one more macroiteration to meet the convergence threshold.  Without enabling the
microiteration of $E_{el}(\mathbf{R})$, it takes 9 and 10 cycles to converge $E$
for electrons coupled with $\nu=0$ and $\nu=1$ vibration, respectively.  The
coupling term $\frac{\grad_\nu\chi(\mathbf{R})}{\chi(\mathbf{R})}$ is
numerically ill-defined at the nodes of $\chi(\mathbf{R})$. Therefore, the PES
$E_{el}(\mathbf{R})$ is generally more difficult to converge for excited nodal
vibrational wavefunction $\chi(\mathbf{R})$.  For instance, the electronic
energy $E_{el}(\mathbf{R})$ for $\nu=1$ level of H$_2$ shows a large numerical
variation and instability at an order of DBOC correction at a bond distance very
close to the node, even the convergence of the molecular energy $E$ is well
achieved.  The significant error in $E_{el}$ has very limited influence on the
convergence of $E$ because nuclear density is negligible at nodes and tails.
The similar convergence pattern is observed for HF molecule.  The $\nu=0$
energies {$E$} and $E_{el}$ are converged within 10$^{-8}$ and $5 \times
10^{-7}$ au in 9 macroiterations. However, for each macroiteration, 3
microiterations are needed on the average. In case of $\nu=1$, up to 17
microiterations are required to converge $E_{el}$ within $10^{-4}$ in the
vicinity of node.  Without using the microiteration, the number of iterative
cycles increases to 24 and 26 for $\nu=0$ and $\nu=1$ vibration, respectively.

\section{\label{sec:res}Results}

\subsection{Molecular energetics}

\begin{figure*}[h]
\subfloat[\label{fig:H2pv0}]{
         \includegraphics[width=0.5\textwidth]{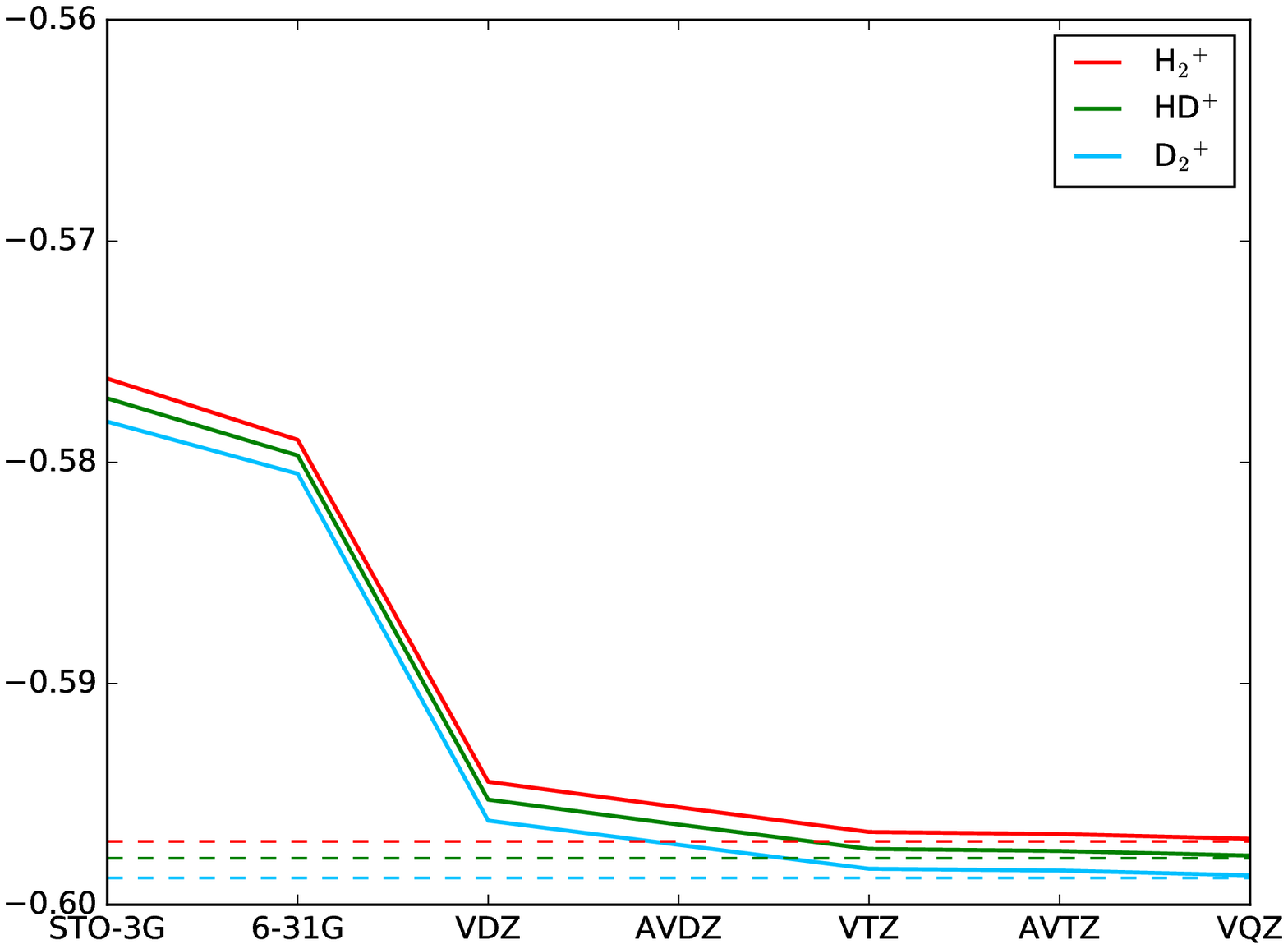}
      } 
\subfloat[\label{fig:H2pv1}]{
         \includegraphics[width=0.5\textwidth]{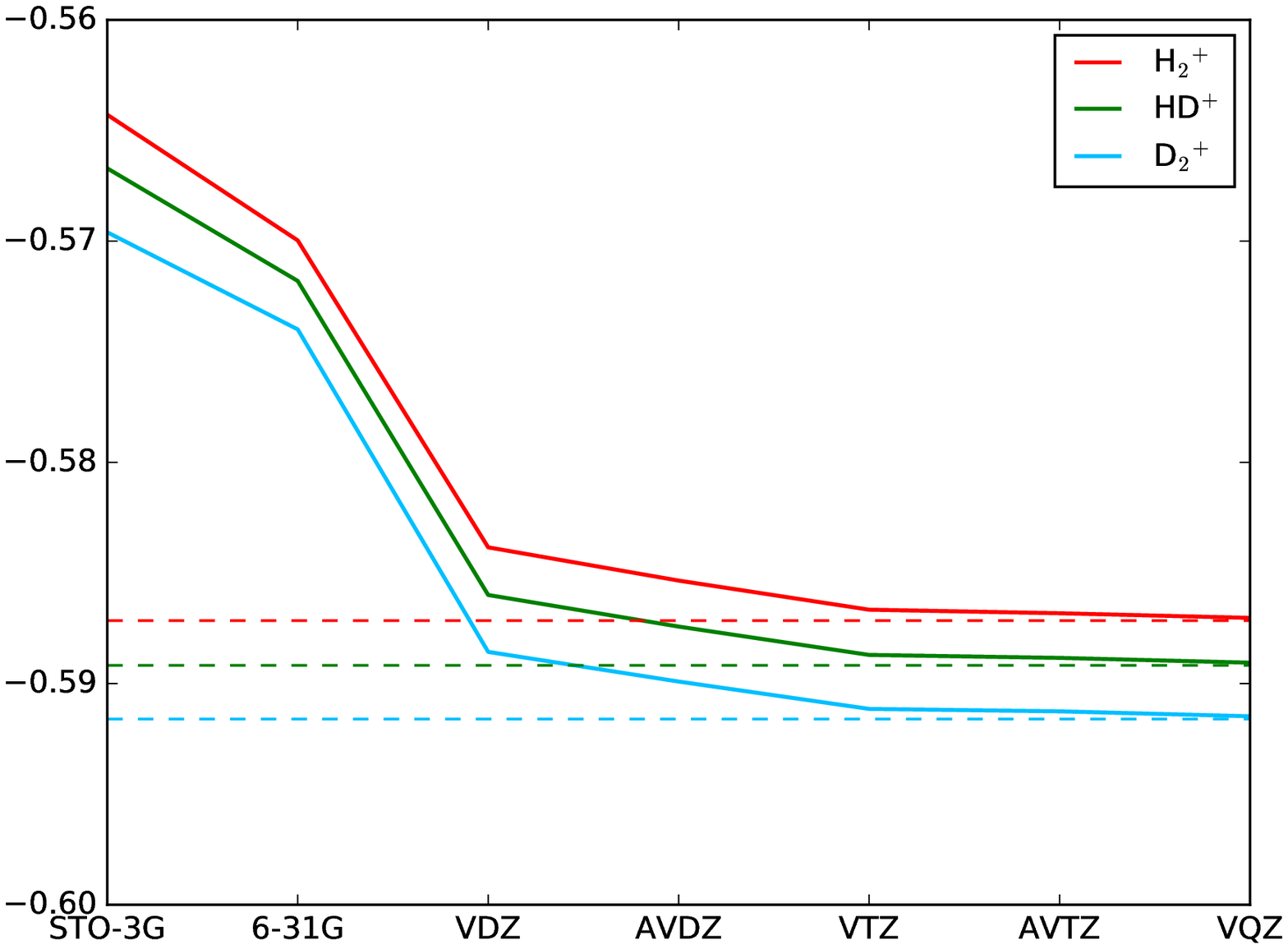}
      } 
\caption{Comparison of the binding energy (au) convergence for
H\textsubscript{2}\textsuperscript{+}, HD\textsuperscript{+} and
D\textsubscript{2}\textsuperscript{+} at $\nu$=0  \textbf{(a)} and $\nu$=1
\textbf{(b)} vibrational level using a range of basis sets. The reference data
from nonadiabatic variational results~\cite{Hilico2000} are
given in dashed lines.}
        \label{fig:H2pConv}
\end{figure*}

The NECSCF-based binding energies of H\textsubscript{2} molecules
(H\textsubscript{2}, HD and D\textsubscript{2}) and their cations
(H\textsubscript{2}\textsuperscript{+}, HD\textsuperscript{+} and
D\textsubscript{2}\textsuperscript{+}) were obtained from the eigenvalue
difference of the nuclear equation Eq.~(\ref{eq:working2})
governing nuclear motion, as shown in
Fig.~\ref{fig:H2pConv} 
for various basis sets. For cations, highly
accurate reference values have been reported in studies using nonadiabatic
variational\cite{Bishop1974,Hilico2000} and QMC~\cite{Alexander2012} methods
which benchmark non-BO energy corrections for these molecules exhibiting no
electronic correlation. 
The NECSCF results of VXZ (X=T, Q, 5) basis sets were 
extrapolated\cite{Helgaker1997,Pansini2016cbs}
to obtain the complete basis set (CBS) limits, as
given in Table~\ref{tab:CBSlimit}.
The details of CBS extrapolation can 
be found in Table S2 and S3 of the Supplementary Materials.
\begin{table*}[h]
\caption{\label{tab:CBSlimit}Comparison of the CBS limits between the BO and non-BO FCI binding
energies ($\Delta E$, au).  The NECSCF nucleus-electron couplings with $\nu=0$
and $\nu=1$ vibrational levels were included for
H\textsubscript{2}\textsuperscript{+}, HD\textsuperscript{+},
D\textsubscript{2}\textsuperscript{+} and H\textsubscript{2}. The reference data
from the nonadiabatic variational method are provided for
comparison~\cite{Hilico2000}.}
\begin{ruledtabular}
\begin{tabular}{lcccc}
Binding energy ($\Delta E$) & H\textsubscript{2}\textsuperscript{+} &   HD\textsuperscript{+}
& D\textsubscript{2}\textsuperscript{+} &H\textsubscript{2} \\
\hline 
\multicolumn{5}{l}{$\nu=0$} \\ [0.5em]
 BO-FCI     & -0.59739883 & -0.59809351 & -0.59891992 & -1.16452199 \\
 NECSCF-FCI & -0.59714185 & -0.59790031 & -0.59879051 & -1.16406555 \\
 Ref.       & -0.59713906\footnotemark[3]& -0.59789797\footnotemark[3] & -0.59878878\footnotemark[3]& -1.16402502\footnotemark[4]  \\
 non-BO correction\footnotemark[1] & 0.00025698 & 0.00019320 & 0.00012941 & 0.00045644 \\
 non-BO error\footnotemark[2] & 0.00000279 & 0.00000234 & 0.00000173 & 0.00004053 \\
\hline
\multicolumn{5}{l}{$\nu=1$} \\ [0.5em]
 BO-FCI     & -0.58741083 & -0.58937451 & -0.59173259 & -1.14555622 \\
 NECSCF-FCI & -0.58715734 & -0.58918220 & -0.59160316 & -1.14510815 \\
 Ref.       & -0.58715568 & -0.58918183 & -0.59160312 & -1.14506537\footnotemark[5] \\
 non-BO correction\footnotemark[1]& 0.00025349 & 0.00019231   & 0.00012943 & 0.00044807\\
 non-BO error\footnotemark[2]& 0.00000166 & 0.00000037 & 0.00000004 & 0.00004278 \\
\end{tabular} 
\end{ruledtabular}
\footnotetext[1]{$\abs{\Delta E_\mathrm{NECSCF-FCI}-\Delta E_\mathrm{BO-FCI}}$.}
\footnotetext[2]{$\abs{\Delta E_\mathrm{NECSCF-FCI}-\Delta E_\mathrm{ref.}}$.}
\footnotetext[3]{From ref.~\citenum{Hilico2000}.}
\footnotetext[4]{From ref.~\citenum{Wolniewicz1995}.}
\footnotetext[5]{From ref.~\citenum{Bubin2009nonadiabatic}.}
\end{table*}
The absolute errors of the uncorrelated
NECSCF binding energies are in an order of
{10\textsuperscript{-6}-10\textsuperscript{-8}} au, indicating 
{a promising} agreement to the benchmark data. 
The non-BO energy corrections to the
conventional BO values are around $10^{-4}$ au, which are {2--3} orders
of magnitude greater than the energy deviations from the benchmark.

For correlated H$_2$ molecule, the electronic correlation is of essential
significance. The CBS limits at NECSCF, NECSCF-MP2 and NECSCF-FCI level of theory
are {-1.12284883, -1.15726946 
and -1.16406555} au, respectively. Obviously,
the electronic correlation correction at the FCI level is a prerequisite to the
successful recovery of the benchmark value $-1.16402502$ au\cite{Wolniewicz1995}
from which our NECSCF-FCI deviates by only {0.04} mau, an order of magnitude
smaller than the nucleus-electron correction of 0.46 mau to the binding energy.
The NECSCF-MP2 ($\nu$=0) and NECSCF-FCI ($\nu$=0) electron correlation energy is
{-0.03442063 and -0.04121672} au in the presence of the explicit nucleus-electron
coupling, as compared to the BO-MP2 and BO-FCI electron correlations of
{-0.03441996 and -0.04121423} au, respectively.  As such, for H$_2$ molecule, the
non-BO corrections to electron correlation are only about ${0.2}~\mathrm{cm}^{-1}$
and ${-0.5}~\mathrm{cm}^{-1}$ from NECSCF-MP2 and NECSCF-FCI correlated models,
respectively, which makes a negligible contribution to the overall non-BO
correction that must be dominated by the NECSCF nucleus-electron coupling among
mean-field electrons.

%In order to verify the open-shell version of the NECSCF solver, the dissociation curves for
%H\textsubscript{2} are computed and compared with the UHF based PES as shown in
%Fig.~\ref{fig:UHFPES}. It can be seen that all curves approach the correct
%dissociation limit at long inter-atomic distances. UHF does not distinguish
%isotopes and therefore gives the same dissociation curve for
%H\textsubscript{2}/HD/D\textsubscript{2}. In contrast, the geometric isotope
%effect can be captured by the NECSCF as a result of the mass dependence in the
%nucleus-electron coupling and nuclear kinetic terms. Because the energy
%correction is positive and the proton is lighter the deuterium, the  non-BO
%electronic energy of H\textsubscript{2} is the most high-lying one at the
%equilibrium position.

Next, we assess the non-BO energy corrections from NECSCF and NECSCF-MP2
computations for many-electron hydride molecules (Table~\ref{tab:EcorrHydride})
by comparing to reported FN-DMC\cite{yang2015large} results. Our NECSCF-MP2
energy corrections for most hydrides deviate from FN-DMC reference within 0.5
mau (1.3 kJ/mol), and the inclusion of MP2 correlations largely decreases the
energy error. 
{The discrepancy of NECSCF corrections from the additive correction
by the zero-point energy (ZPE) and DBOC can be viewed as contribution beyond the
limit of BO electronic states, including nucleus-electron correlation effects on
modifying electronic potential, ZPE and DBOC. The remaining non-BO correction is
$>$ 0.2 mau for BH, CH ($^4\Sigma$) and HF.}
In general, it is noted that most of the nucleus-electron couplings are promisingly 
captured at the mean-field NECSCF level.  However, a relatively large discrepancies 
of {2.4 mau (6.3 kJ/mol)} and 1.8 mau (4.7 kJ/mol)
occur to the doublet CH ($^2\Pi$) and quartet CH ($^4\Sigma$) states,
respectively, probably  due to their strong multi-reference characters for which
the electronic correlation at MP2 level is insufficient. The implementation of
NECSCF-based multi-state multi-reference correlation methods such as
NECSCF-CASSCF and NECSCF-NEVPT2 (NECSCF-based N-electron Valence State
Perturbation Theory) will be reported in future.
\begin{table}[h]
\caption{Non-BO energy corrections (mau) to first-row hydrides.
 \label{tab:EcorrHydride}}
\begin{ruledtabular}
\begin{tabular}{lcccc}
 \textbf{Molecule} & \textbf{NECSCF} & \textbf{NECSCF-MP2} & \textbf{FN-DMC}\footnotemark[1]  & \textbf{ZPE+DBOC}\footnotemark[1] \\
\hline
LiH  & 4.09(6) & 4.14(5) & 4.28(3) & 4.07(2) \\
BeH  & 5.71(6) & 5.76(0) & 5.99(6) & 5.90(1) \\
BH   & 7.25(0) & 7.29(1) & 7.39(9) & 7.03(2) \\
CH ($^2\Pi$)   & 8.50(4) & 8.37(5) & \multirow{2}{*}{10.8(3)} & \multirow{2}{*}{8.54(9)} \\
CH ($^4\Sigma^-$) & 9.29(4) & 9.03(9) & & \\
OH   & 11.5(7) & 11.1(2) & 11.1(5) & 11.1(0) \\
HF   & 12.8(8) & 12.3(7) & 12.0(4) & 12.1(4) 
\end{tabular}
\end{ruledtabular}
\footnotetext[1]{From ref.~\citenum{yang2015large}}
\end{table}

The NECSCF and NECSCF-MP2 proton affinities (PAs) of five diatomic molecules
were calculated using the procedure described in
Ref.~\citenum{pavosevic2020multicomponent}, and are compared with the
experimental\cite{Hunter1998}, NEO-HF and
NEO-OOMP2\cite{pavosevic2020multicomponent} values. For the calculation of
protonated species, all vibrational modes (3 for non-linear and 4 for linear
triatomic molecules) were selected to couple individually with electronic MOs,
whereas only the results with the modes yielding the smallest absolute error from
experimental data are presented in Table~\ref{tab:ProtonAffinity}. 
The PAs associated with the remaining molecular vibrations are provided in Table
S1 of the Supplementary Materials.  NECSCF-based methods achieve the best
performance for the asymmetric proton stretching of H-XY molecule, where the
proton departure from XY\textsuperscript{-} decreases the X-Y bond length.  This
is consistent with the relatively shorter equilibrium bond distance in
XY\textsuperscript{-} than in H-XY molecule. The absolute error of NECSCF is
generally smaller than NEO-HF since the nuclear anharmonic motion and ZPE 
are exactly incorporated through FGH solution. NECSCF-MP2 is also
able to achieve a comparable accuracy to the NEO-OOMP2 values.
\begin{table*}[h]
\caption{Absolute deviations (eV) of the PAs from experimental data.
\label{tab:ProtonAffinity}}
\begin{ruledtabular}
\begin{tabular}{lccccc}
 \textbf{Molecule} & \textbf{Experiment}\footnotemark[1] &\textbf{NECSCF} & \textbf{NEO-HF}\footnotemark[2] & \textbf{NECSCF-MP2}  &
\textbf{NEO-OOMP2}\footnotemark[2] \\
\hline
CN\textsuperscript{-}   & 15.31 & 0.20 & 0.91 & 0.34 & 0.29 \\
N\textsubscript{2}      & 5.12  & 0.00 & 0.76 & 0.03 & 0.23 \\
HS\textsuperscript{-}   & 15.31 & 0.25 & 0.84 & 0.29 & 0.31 \\
OH\textsuperscript{-}   & 16.95 & 0.50 & 0.36 & 0.07 & 0.42 \\
CO                      & 6.16  & 0.02 & 0.84 & 0.10 & 0.04 
\end{tabular}
\end{ruledtabular}
\footnotetext[1]{From ref.~\citenum{Hunter1998}.}
\footnotetext[2]{From ref.~\citenum{pavosevic2020multicomponent}.}
\end{table*}

\subsection{Bond length}

BO quantum chemistry methods associate equilibrium molecular structures with the
energy minima on the PES of molecule. However, the presence of ZPE, anharmonic
vibrations and non-BO couplings necessitates the determination of vibrationally
averaged properties to reflect the thermal nature of atomic positions.  
\cite{Costain1958,Hargittai1992}
Here, we demonstrate the NECSCF computation of vibrationally averaged bond
lengths $\expval{R}$ and the non-BO impact for typical diatomic and
triatomic molecules,
according to the expectation value of the single product molecular wavefunction
fulfilling PNC condition.
\begin{equation}
\expval{R} = \mel{\Psi(\mathbf{rs},\mathbf{R})}{\mathbf{R}}
{\Psi(\mathbf{rs},\mathbf{R})}_\mathbf{rsR} = \mel{\chi(\mathbf{R})}
{\mathbf{R}}{\chi(\mathbf{R})}_\mathbf{R}
\end{equation}
The NECSCF and NECSCF-MP2 results averaged on the ground $\nu=0$ and first
excited $\nu=1$ vibrational modes are compared with the full-quantum
cNEO-DFT\cite{Xu2020} and experimental values in Table~\ref{tab:BondLength}.

\begin{figure}[h]
     \includegraphics[width=0.5\textwidth]{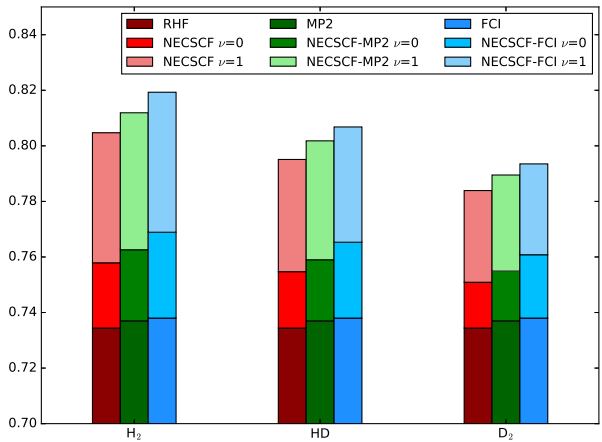}
     \caption{Vibrationally averaged bond lengths (\AA) of H\textsubscript{2}.}
     \label{fig:rHH}
\end{figure}

For all computed chemical bonds involving H
atom, it can be seen that the non-BO nucleus-electron coupling effects lead to
a notable increase of equilibrium bond length. 
In contrast, for
the bonds between heavier atoms, the non-BO coupling with $\nu=0$ vibration
leads to the bond length elongation
of only $\sim 0.01$ {\AA} for D-F, D-C, D-N and D-O bonds and $<0.005$ {\AA}
for C-N bonds, respectively, as compared to both BO RHF and MP2 results. 
The $\nu=1$ vibration results in further expansion of
atomic positions and longer averaged bond lengths than those with $\nu=0$.
It is also observed that deuteration shortens both NECSCF and NECSCF-MP2 bond
lengths. For instance, the H-F bond length is larger than D-F by $0.004$ {\AA}.
\begin{table*}
\caption{Vibrationally averaged bond lengths ({\AA}) of selected molecules at
$\nu=0$ and $\nu=1$ vibrational levels. For computed bond lengths of triatomic
molecules, only the bond stretching mode of interest is active with other modes
frozen. For comparisons, the regular equilibrium RHF and MP2 bond lengths are
equivalent to vibrational averages in the BO harmonic potential; the
experimental bond lengths were determined by fitting the BO spectroscopic
constants which lead to the same geometric parameters for all isotopomers. 
}
\label{tab:BondLength}
\footnotesize
\begin{ruledtabular}
\begin{tabular}{lccccccccc}
\multirow{2}{*}{\textbf{Molecule}} & \multirow{2}{*}{\textbf{Bond}} &
\multirow{2}{*}{\textbf{Experiment}\footnotemark[1]} &
\multirow{2}{*}{\textbf{RHF}} & \multicolumn{2}{c}{\textbf{NECSCF}} &
\multirow{2}{*}{\textbf{MP2}} & \multicolumn{2}{c}{\textbf{NECSCF-MP2}} &
\multirow{2}{*}{\textbf{cNEO-DFT}\footnotemark[2]} \\ 
    &     &  & & \textbf{$\nu=0$} & \textbf{$\nu=1$} & & \textbf{$\nu=0$} & \textbf{$\nu=1$} & \\
\hline
    H\textsubscript{2}  & H-H &  0.7414 & 0.7344 & 0.7579 & 0.8047 & 0.7370 & 0.7626 & 0.8119 & 0.785 \\
    HD                  & H-D &	 0.7414 &        & 0.7547 & 0.7951 &        & 0.7590 & 0.8018 & 0.779 \\
    D\textsubscript{2}  & D-D &	 0.7415 &        & 0.7509 & 0.7839 &        & 0.7549 & 0.7895 & 0.760 \\
    HF			& H-F &	 0.9168 & 0.8979 & 0.9121 & 0.9371 & 0.9169 & 0.9318 & 0.9528 & 0.942 \\
    DF 			& D-F &	        &        & 0.9083 & 0.9292 &        & 0.9281 & 0.9501 & 0.937 \\
\hline
    HCN			& C-H &  1.064	& 1.0570 & 1.0709 & 1.0954 & 1.0599 & 1.0737 & 1.0980 & 1.089 \\
                        & C-N &  1.156 	& 1.1337 & 1.1369 & 1.1431 & 1.1808 & 1.1850 & 1.1922 & 1.147 \\
    DCN 		& D-C & 	&        & 1.0671 & 1.0869 &        & 1.0697 & 1.0901 & 1.082 \\
			& C-N &         &        & 1.1368 & 1.1429 &        & 1.1850 & 1.1922 & 1.147 \\ 
    HNC			& N-H &  0.986	& 0.9830 & 0.9968 & 1.0223 & 0.9921 & 1.0067 & 1.0316 & 1.019 \\
                        & C-N &  1.173 	& 1.1558 & 1.1592 & 1.1660 & 1.1909 & 1.1951 & 1.2026 & 1.165 \\
    DNC 		& D-N & 	&        & 0.9930 & 1.0128 &        & 1.0027 & 1.0226 & 1.012 \\
			& C-N &         &        & 1.1592 & 1.1659 &        & 1.1950 & 1.2025 & 1.165 \\ 
\hline
    H\textsubscript{2}O	& H-O &  0.958  & 0.9406 & 0.9552 & 0.9847 & 0.9560 & 0.9716 & 1.0031 & 0.981 \\ 
    HDO                 & H-O &  0.956  &        & 0.9552 & 0.9846 &        & 0.9716 & 1.0030 & 0.982 \\ 
			& D-O &  0.956  &        & 0.9512 & 0.9725 &        & 0.9672 & 0.9899 & 0.975 \\ 
    D\textsubscript{2}O	& D-O &  0.956  &        & 0.9511 & 0.9723 &        & 0.9672 & 0.9898 & 0.975 
\end{tabular} 
\end{ruledtabular}
\footnotetext[1]{From ref.~\citenum{NIST}.}
\footnotetext[2]{From ref.~\citenum{Xu2020}.}
\end{table*}
The isotopic effect influences the non-BO bond length by both electronic and
vibrational factors: on one hand, the significance of the nucleus-electron
coupling for electronic motion decreases with increasing atomic mass.  On the other
hand, large atomic mass tends to narrow the landscape of nuclear wavefunction,
which leads to a decrease of averaged bond length. 
Finally, the average bond lengths at the NECSCF-FCI level were further computed for
H\textsubscript{2} molecules. The comparison with NECSCF and NECSCF-MP2 are
presented in Fig.~\ref{fig:rHH}. 
Among the hydrogen molecules,
H\textsubscript{2} has the largest bond length increase, whereas the change of
D-D interatomic distance is the smallest.

\subsection{Nuclear and electron density}

\begin{figure*}[h]
\subfloat[\label{fig:NucDenH2}]{
         \includegraphics[width=0.5\textwidth]{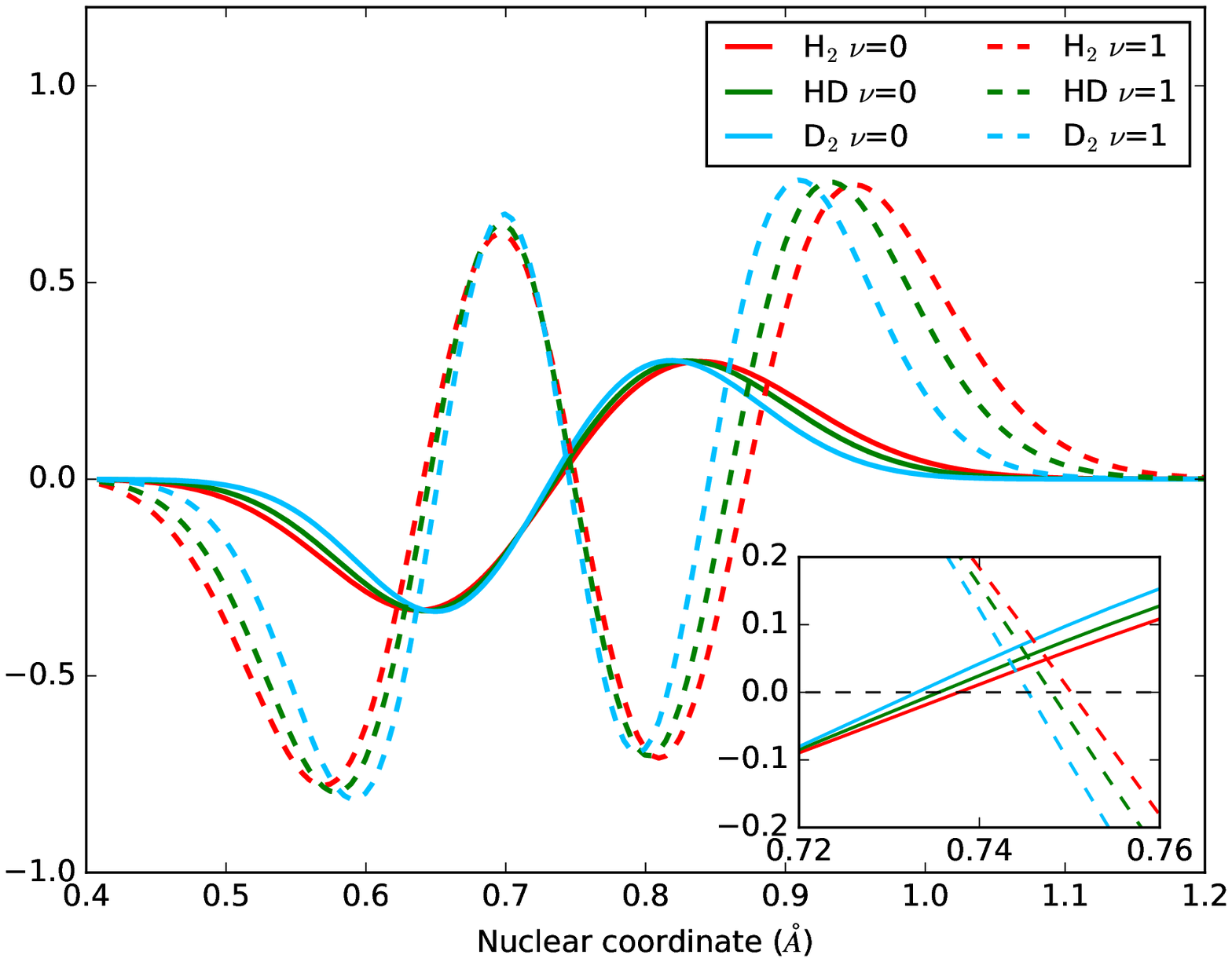}
         }
\subfloat[\label{fig:NucDenHF}]{
         \includegraphics[width=0.5\textwidth]{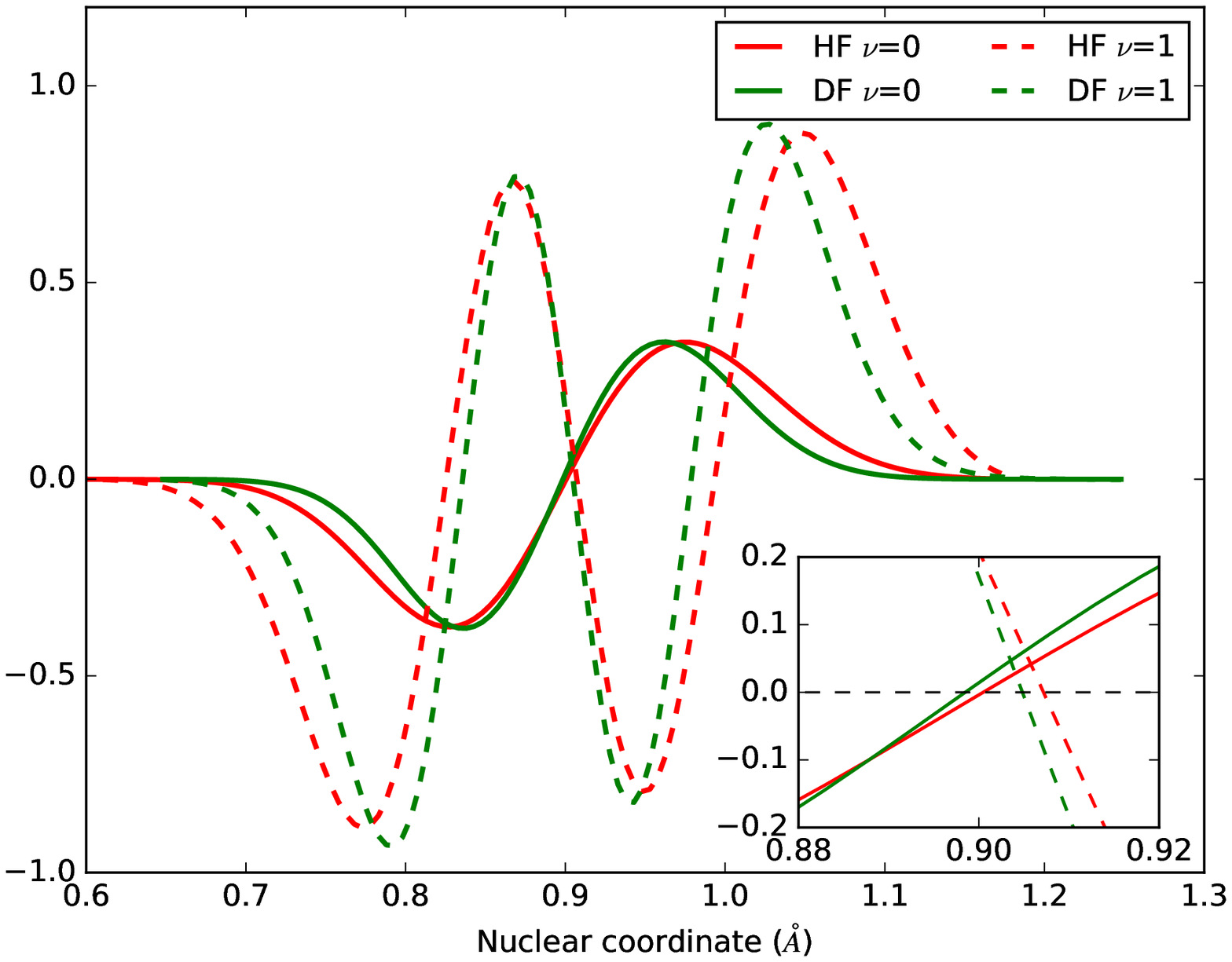}
         }
	\caption{The NECSCF correction (au) to {quantum HO} nuclear density of 
H\textsubscript{2} \textbf{(a)} and HF \textbf{(b)}. The inset shows the details around the
equilibrium bond lengths.
}
        \label{fig:NucDen}
\end{figure*}

In this section, we exam the non-BO effects on the nuclear and electronic
density distribution. At our disposal, the NECSCF method discards 
molecular translational and rotational motions. The nuclear density
$P_\nu(\mathbf{R})$ only depends on the interatomic distance
$\mathbf{R}$ of diatomic molecules and the excitation level $\nu$ of the nuclear wavefunction $\chi_\nu$.
Here, we define the non-BO correction to nuclear density $\Delta P_\nu(\mathbf{R})$ as 
\begin{equation}
\Delta P_\nu(\mathbf{R}) = P^{\mathrm{NECSCF}}_\nu(\mathbf{R})-P^{\mathrm{HO}}_\nu(\mathbf{R})
\end{equation}
with 
\begin{eqnarray}
P^{\mathrm{NECSCF}}_\nu(\mathbf{R}) &= &
\chi_{\nu,\text{NECSCF}}^\dagger(\mathbf{R})\chi_{\nu,\text{NECSCF}}(\mathbf{R}), \\
P^{\mathrm{HO}}_\nu(\mathbf{R}) &=& \chi_{\nu,\text{HO}}^\dagger(\mathbf{R})\chi_{\nu,\text{HO}}(\mathbf{R})
\end{eqnarray}
where $\chi_{\nu,\text{NECSCF}}$ is the nuclear probability amplitude calculated
from NECSCF PES and $\chi_{\nu,\text{HO}}$ the quantum harmonic oscillator 
{(HO)}
function for the $\nu$-th vibrational level.  The non-BO density corrections for
two low-lying states [$\Delta P_0(\mathbf{R})$ and $\Delta P_1(\mathbf{R})$] for
H\textsubscript{2}, HF and their isotopes are graphed in
Fig.~\ref{fig:NucDen}.
Specifically, the non-BO correction $\Delta P_{\nu=0}(\mathbf{R})$ transfers the
nuclear density from the short bond region to the long bond region, e.g., from
0.4--0.7 {\AA} to 0.7--1.2 {\AA} for H\textsubscript{2}. This actually reveals
an increase of the vibrationally averaged bond length for both
H\textsubscript{2} and HF, which is consistent with the results presented in
previous section.  $\Delta P_{\nu=1}(\mathbf{R})$ is apparently much stronger than
$\Delta P_{\nu=0}(\mathbf{R})$, which is also in line with the bond elongation as
shown in Fig.~\ref{fig:rHH}. 
Moreover, the deuteration, which involves heavier atoms and diminishes the
nucleus-electron coupling,  shifts the nuclear density more significantly
towards the equilibrium bond region.  This results in less bond elongation with
more deuteration, following the same isotopic effect as the averaged bond
distance in the order of D\textsubscript{2} $<$ HD $<$ H\textsubscript{2} and DF $<$
HF.

\begin{figure*}[h]
\subfloat[\label{fig:DyDenH2}]{
\includegraphics[width=0.5\textwidth]{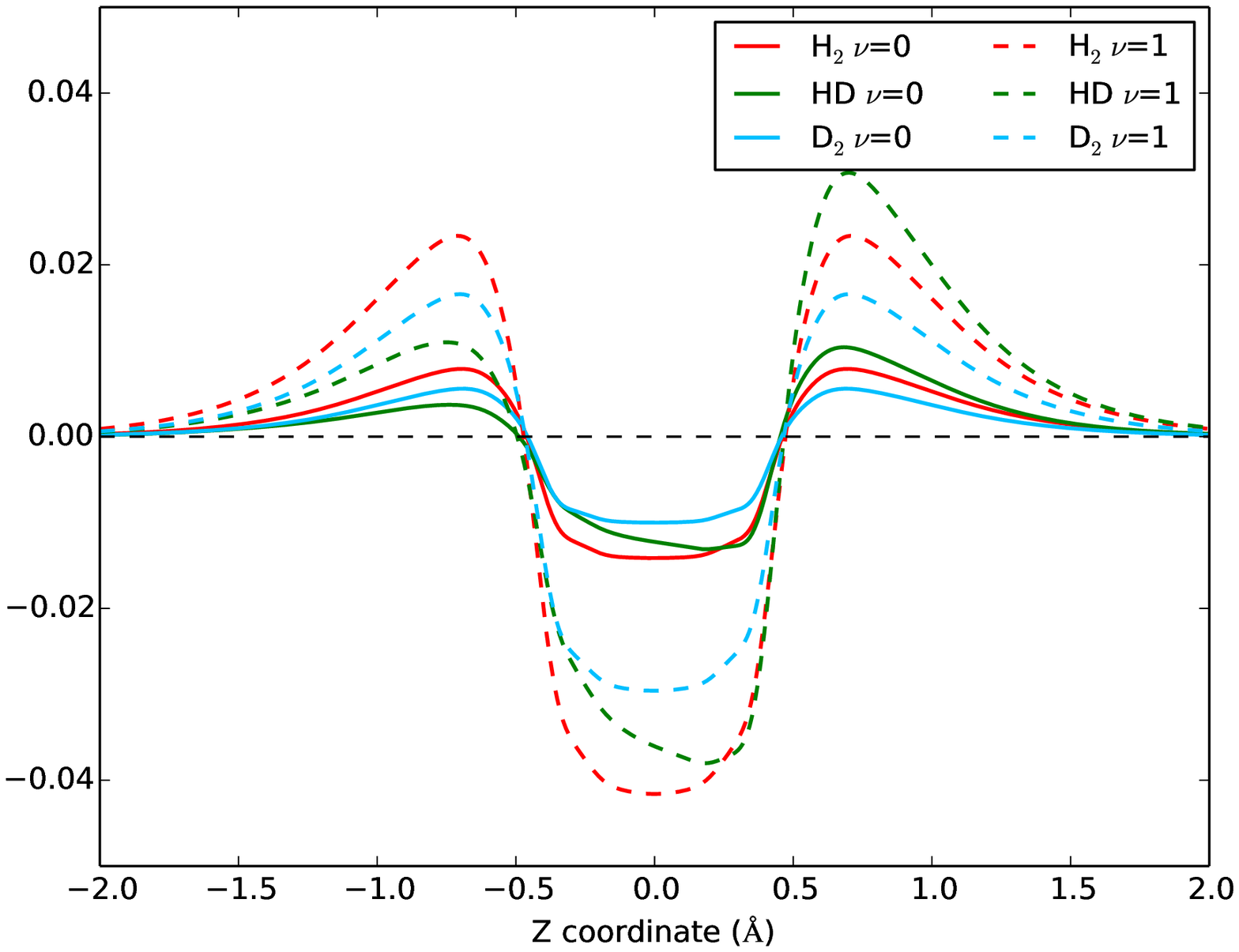}
}
\subfloat[\label{fig:DyDenHF}]{
\includegraphics[width=0.5\textwidth]{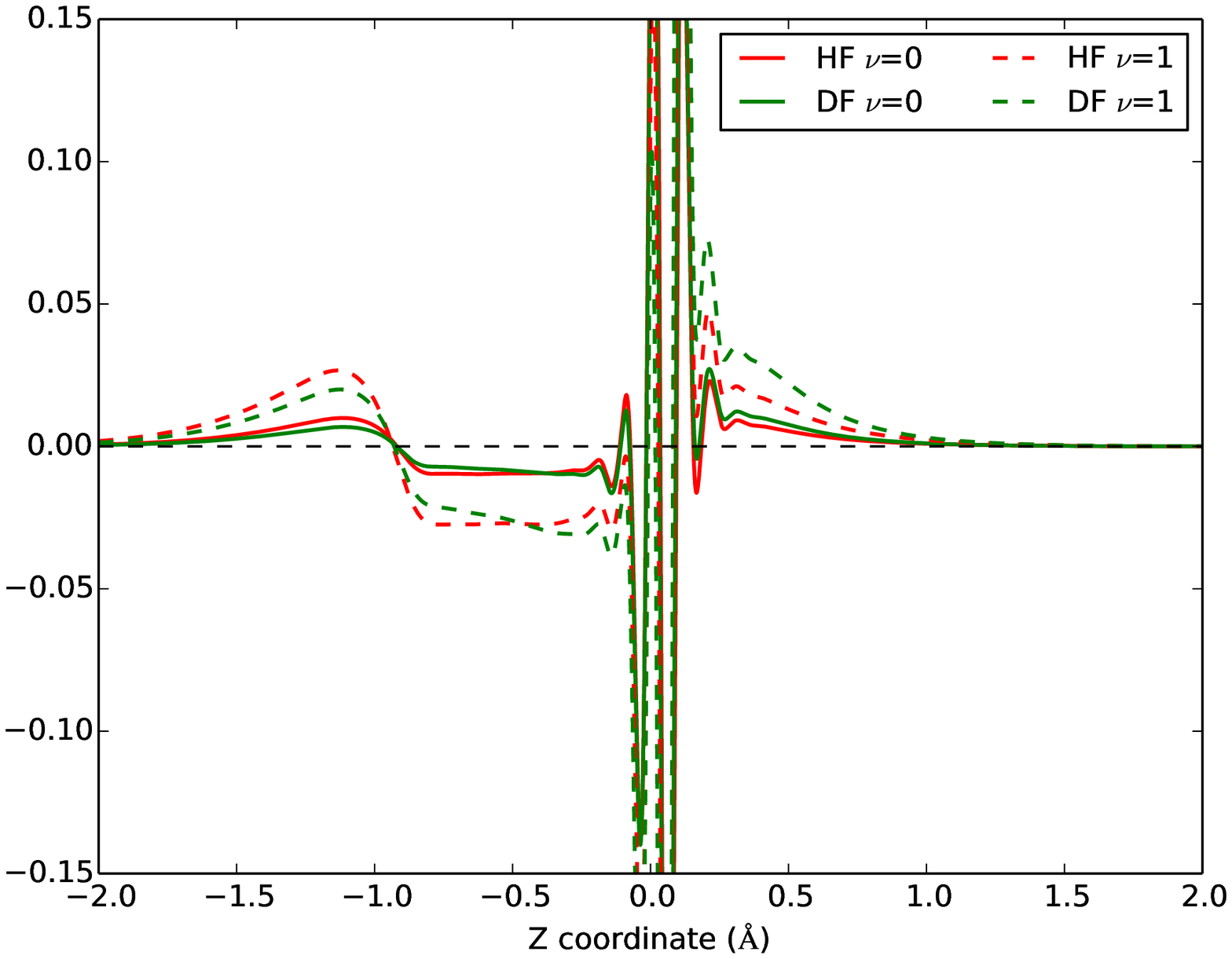}
}
\caption{The $z$-component of dynamic electron density corrections (au)
of H\textsubscript{2} (HD and D$_2$) \textbf{(a)} and HF (DF) \textbf{(b)}. $z$
represents the molecular axis.}
\label{fig:DyDen}
\end{figure*}

Based on the idea of the dynamic distribution of electron density
\cite{Tachibana1984}, we characterize the non-BO correction ($\expval{
\Delta\rho}_{\nu}$) to the dynamic electron density as,  
\begin{equation}
\expval{\Delta\rho}_{\nu} = \expval{\rho}_{\nu,\text{NECSCF}}-\expval{\rho}_{\nu,\text{HO}}
\end{equation}
and the $z$-component of $\expval{\Delta\rho}_{\nu}$ in the direction of
molecular axis
\begin{equation}
\expval{\Delta\rho}_{\nu}^z = \int\int\text{d}x\text{d}y\expval{\rho}_{\nu}.
\end{equation}
Here, $\expval{\rho}_{\nu,\text{NECSCF}} = \int\text{d}\mathbf{R}\rho
P^\mathrm{NECSCF}_{\nu}(\mathbf{R})$ and $\rho$ is the electron density from
NECSCF computation.  $\expval{\Delta\rho}_0^z$ and $\expval{\Delta\rho}_1^z$ for
H\textsubscript{2} are shown in Fig.~\ref{fig:DyDenH2}.

\begin{figure*}[h]
\subfloat[\label{fig:3DDyDenH2}]{
\includegraphics[height=0.25\textheight]{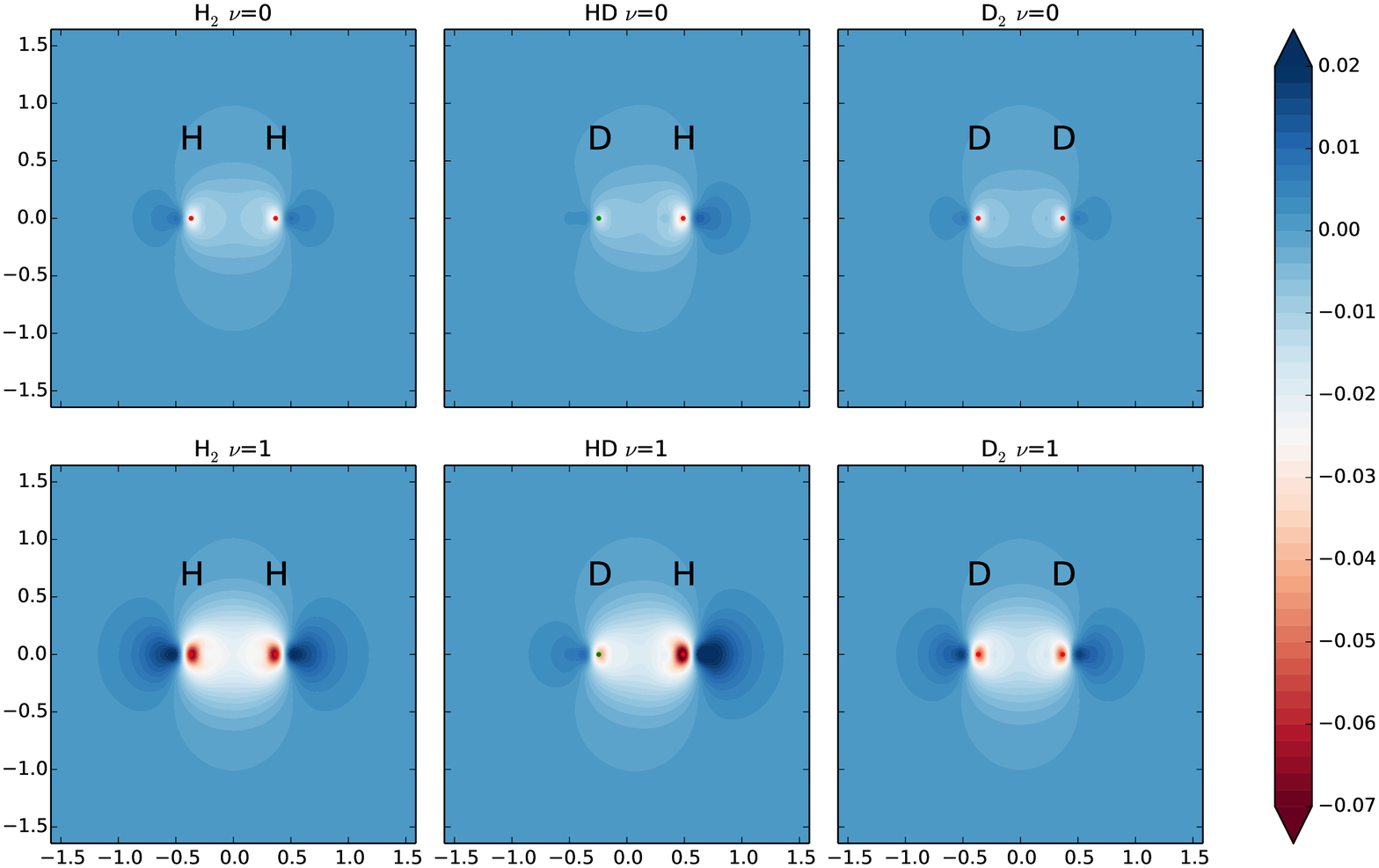}
}
\subfloat[\label{fig:3DDyDenHF}]{
         \includegraphics[height=0.25\textheight]{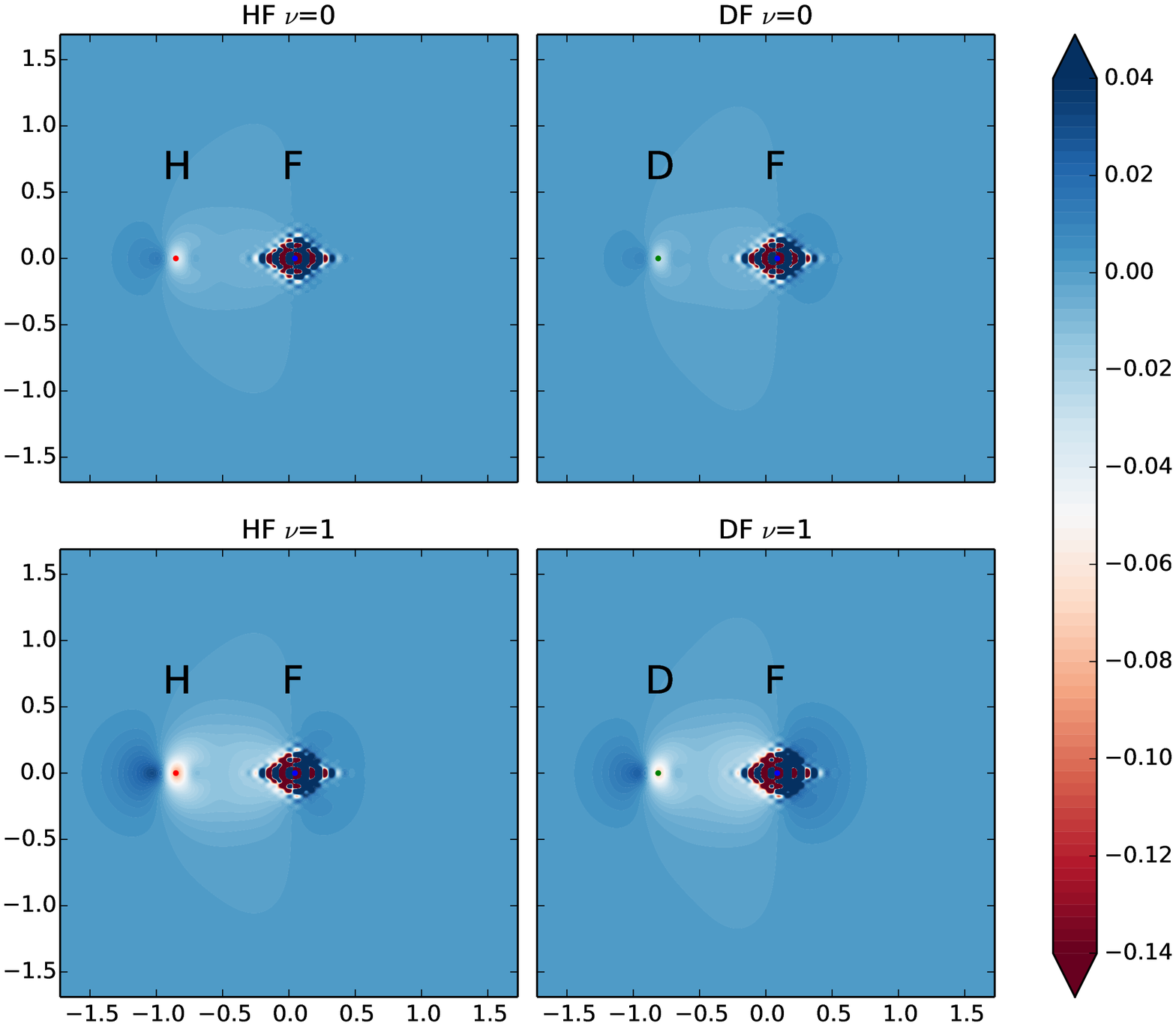}
}
     \caption{The $yz$-contour of dynamic electron density corrections (au) of
H\textsubscript{2} (HD and D\textsubscript{2}) \textbf{(a)} and HF (DF)
\textbf{(b)}.}
        \label{fig:3DDyDen}
\end{figure*}

Upon the nucleus-electron coupling, symmetric density transfer takes place from
the interstitial bonding region between atoms to the peripheral region in
H\textsubscript{2} and D\textsubscript{2}. Due to the decreased nuclear quantum
effect with heavier isotopes, the electron density transfer is obviously less
significant in D\textsubscript{2} than H\textsubscript{2}. For heteronuclear HD,
the dynamic electron transfer becomes asymmetric. Moreover, it is observed that
the transfer of electron density is highly directional from the bonding to the
surrounding area. As illustrated from the $yz$-contour of $\expval{\Delta\rho}$
(Fig.~\ref{fig:3DDyDen}), 
the non-BO construction of
electron density is much stronger along the $z$-direction in which H-H vibrates
than in other orientations, which thus weakens and elongates the $\sigma$-bond.
The $\nu=1$ level of vibration appears to further enhance the density transfer
that follows a similar pattern with that for $\nu=0$ and leads to even longer
bond lengths.
For HF molecule, however, an intensively local variation of electron density is
observed in the near proximity of F atom which possesses strong
electronegativity, and becomes much more local than H\textsubscript{2}.
Nonetheless, more electron density destruction can be still discernible in the F
atom's bonding region than the opposite side.

\subsection{FHF\textsuperscript{-} molecule}

\begin{figure}[h]
\includegraphics[width=0.5\textwidth]{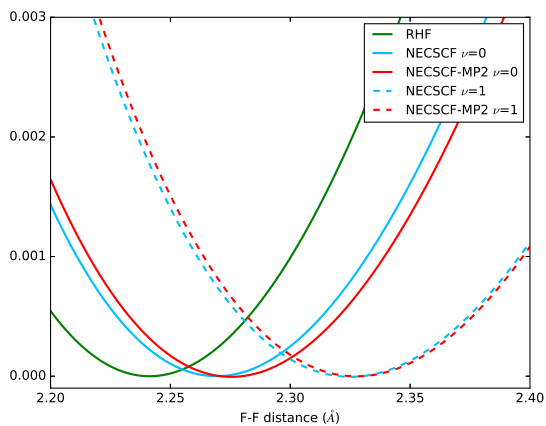}
 \caption{Comparison of the PES as a function of F-F distances for
FHF\textsuperscript{-} between standard RHF, NECSCF and NECSCF-MP2 computations.
The energy minima of all curves are shifted for comparison convenience.}
\label{fig:FHF-PES}
\end{figure}

As an illustrative application of our NECSCF and NECSCF-MP2 implementations, we
demonstrate the nucleus-electron effect on FHF\textsuperscript{-} molecule for
which the paradigmatic proton-electron coupling has been well studied in
literature. The computed NECSCF and NECSCF-MP2 PESs incorporating the
nucleus-electron coupling individually with $\nu=0$ and $\nu=1$ proton shuttle
mode are shown in Fig.~\ref{fig:FHF-PES} 
where the non-BO behavior is compared with standard RHF
PES.  The difference in minimal energy for F-F lengths between NECSCF ($\nu$=0)
and standard RHF was computed to be 0.026 {\AA} which is in excellent agreement
with reported 0.02 {\AA} arising from the proton quantum effect\cite{Hammes2017}.
Interestingly, although the electronic correlation elongates F-F bond, the
change made by MP2 correlation is not as significantly strong as the proton
quantum effect. 

\begin{figure*}[h]
\subfloat[\label{fig:FHF_NucDen0}]{
   \includegraphics[width=0.5\textwidth]{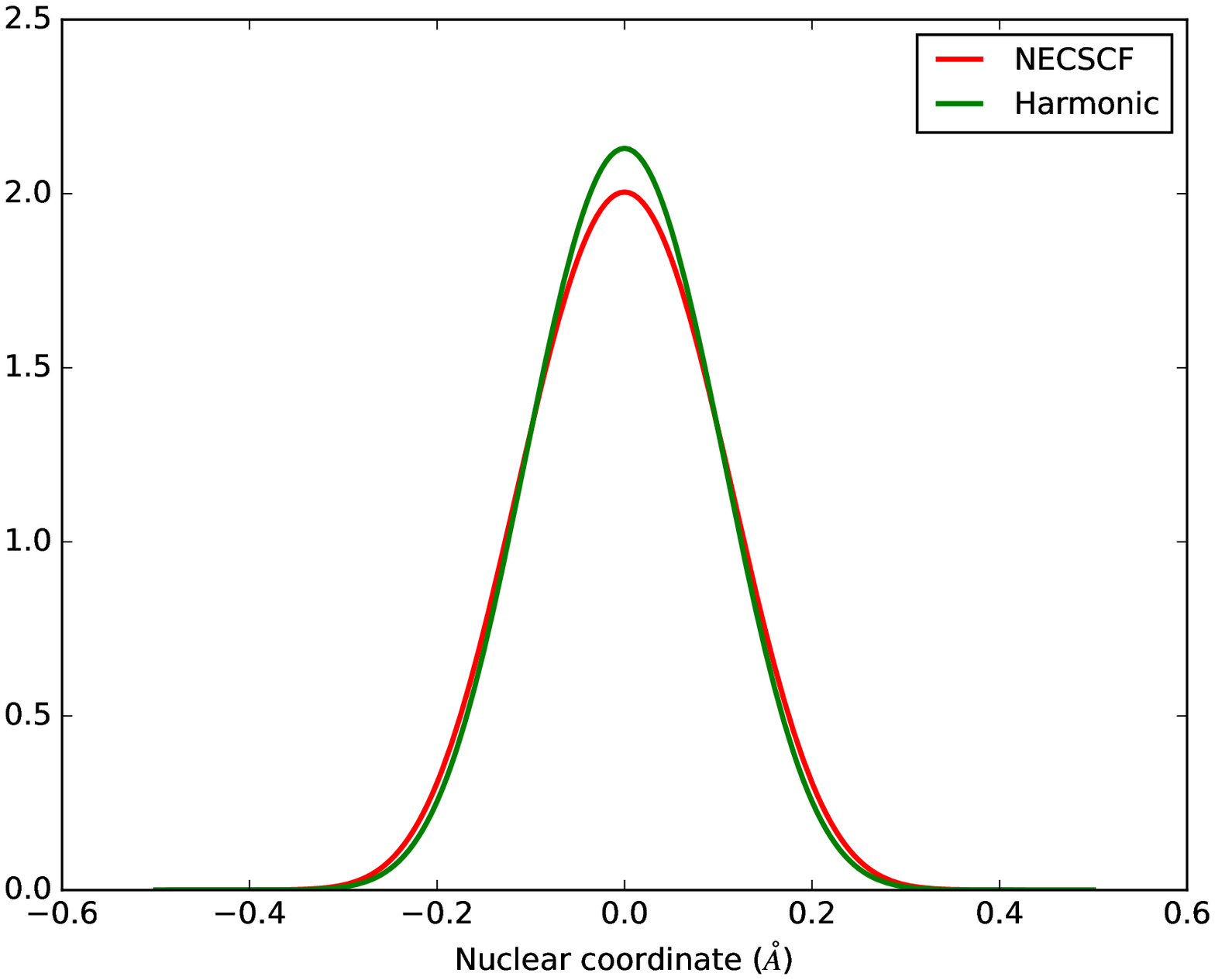}
}
\subfloat[\label{fig:FHF_NucDen1}]{
   \includegraphics[width=0.5\textwidth]{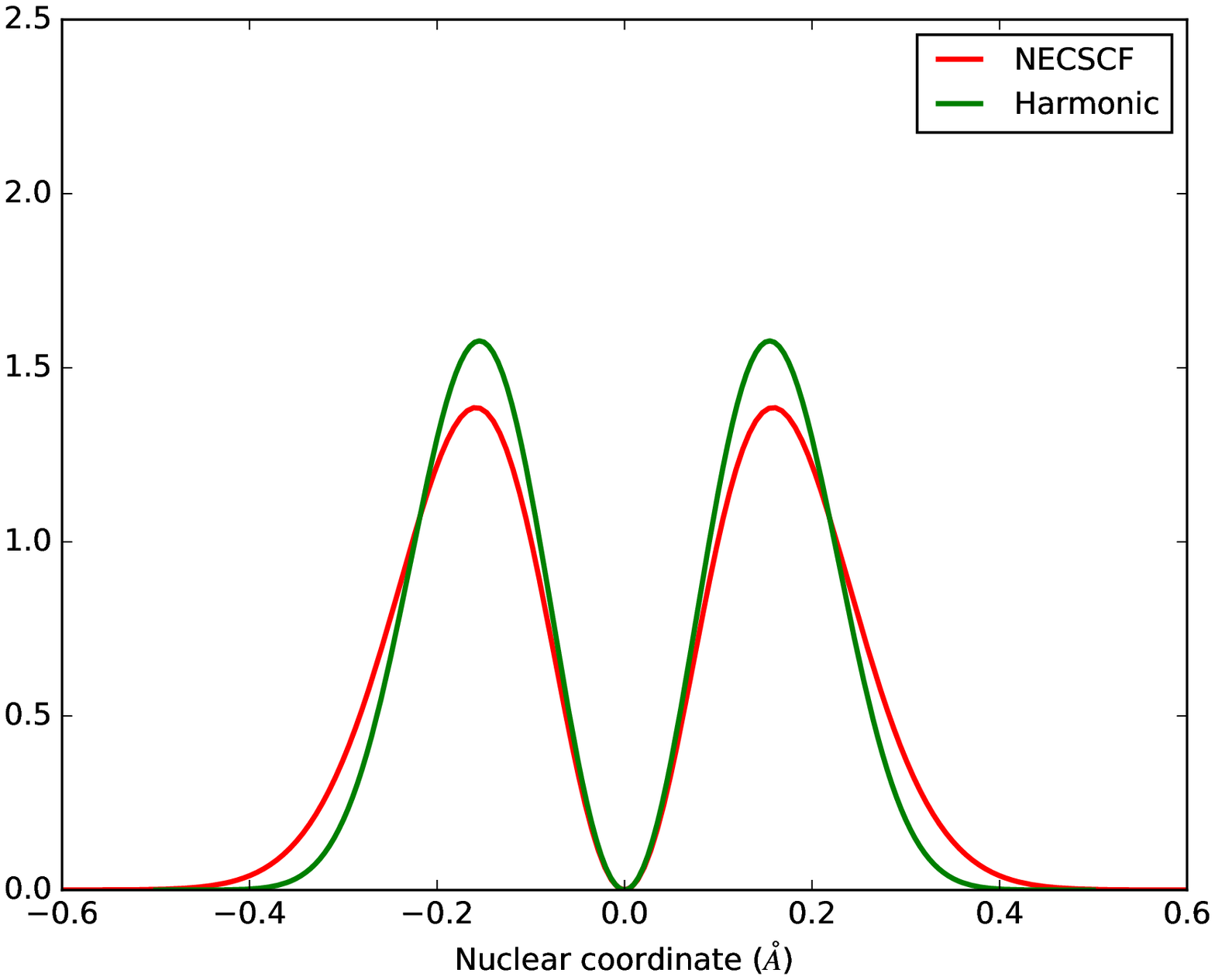}
}
\caption{NECSCF and {harmonic} protonic density (au) of FHF\textsuperscript{-}
for $\nu$=0 \textbf{(a)} and $\nu$=1 \textbf{(b)} vibrational levels. 
{The vibrational frequency of the quantum HO is obtained from 
RHF calculation. }
The F-F distances are determined according to the PES minima for NECSCF and RHF,
respectively.}
        \label{fig:FHF_NucDen}
\end{figure*}

Using the tool we developed in previous section, the nuclear and electron
density analyses for the proton quantum effect are shown in
Fig.~\ref{fig:FHF_NucDen}.  
\begin{figure*}
\subfloat[\label{fig:FHF_DyDen0}]{
\includegraphics[width=0.5\textwidth]{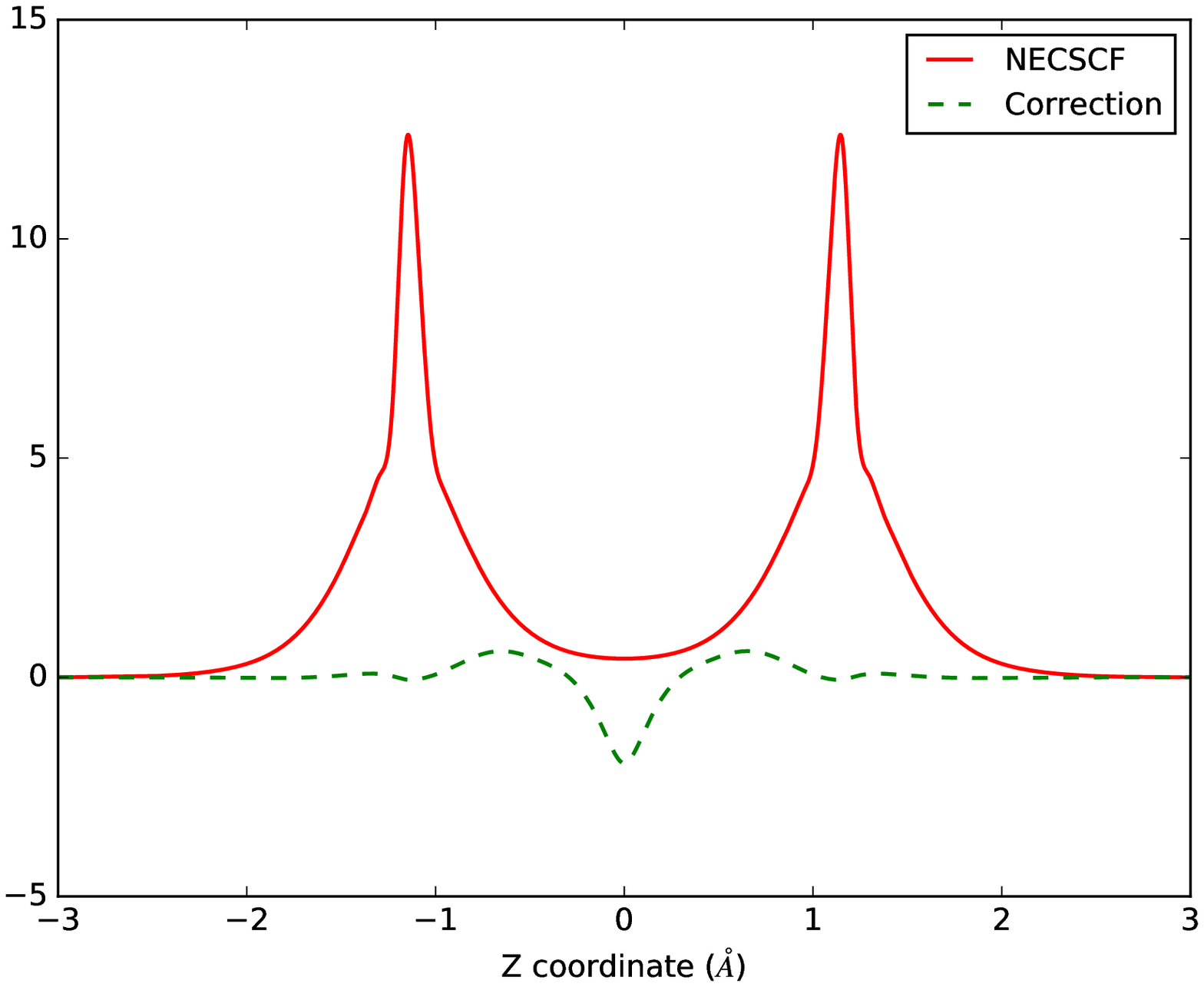}
}
\subfloat[\label{fig:FHF_DyDen1}]{
\includegraphics[width=0.5\textwidth]{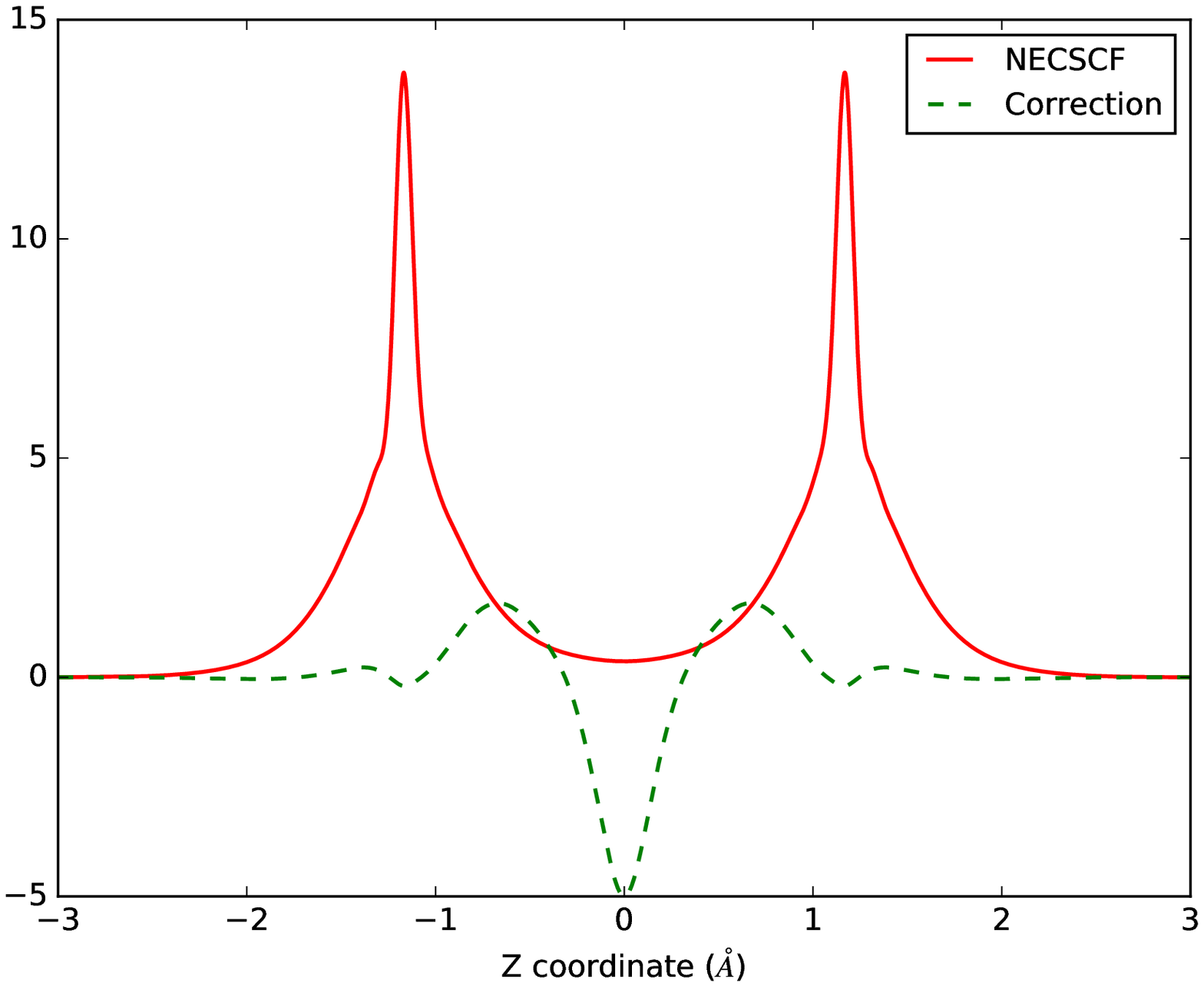}
}
	\caption{The $z$-component of NECSCF dynamic electron density (au) of
FHF\textsuperscript{-} for $\nu$=0  \textbf{(a)} and $\nu$=1 \textbf{(b)}
vibrational levels. The correction to the electron density is magnified by 100
times to exhibit  the protonic non-BO significance.  The F-F distance is
determined according to the NECSCF PES minimum.}
	\label{fig:FHF_DyDen}
\end{figure*} 
In consistence with the NEO-DFT
prediction\cite{Hammes2018}, the majority of the protonic density at $\nu=0$
vibrational state lies in a range of -0.3--0.3 {\AA}. As compared to 
{harmonic result}, the
peak of the NECSCF protonic density decreases with slightly broadened
distribution width, and the tail decay of the NECSCF density is relatively
slower than the RHF one. These subtle changes should be attributed to the
inclusion of the nucleus-electron coupling. The first excited $\nu=1$ protonic
density exhibits changes in a similar fashion with $\nu=0$, but with greater
density broadening.  For electron density distribution (Fig.~\ref{fig:FHF_DyDen}), 
the correction to the dynamic density is rather small, primarily due to the low
electron density on the proton and strongly electronegative F atoms.  The
density correction is symmetric as anticipated, and the density transfers from
the equilibrium proton position towards both left and right H-F bonding regions. 

\begin{figure*}
\subfloat[\label{fig:FHF_3DNucDen0}]{
   \includegraphics[width=0.5\textwidth]{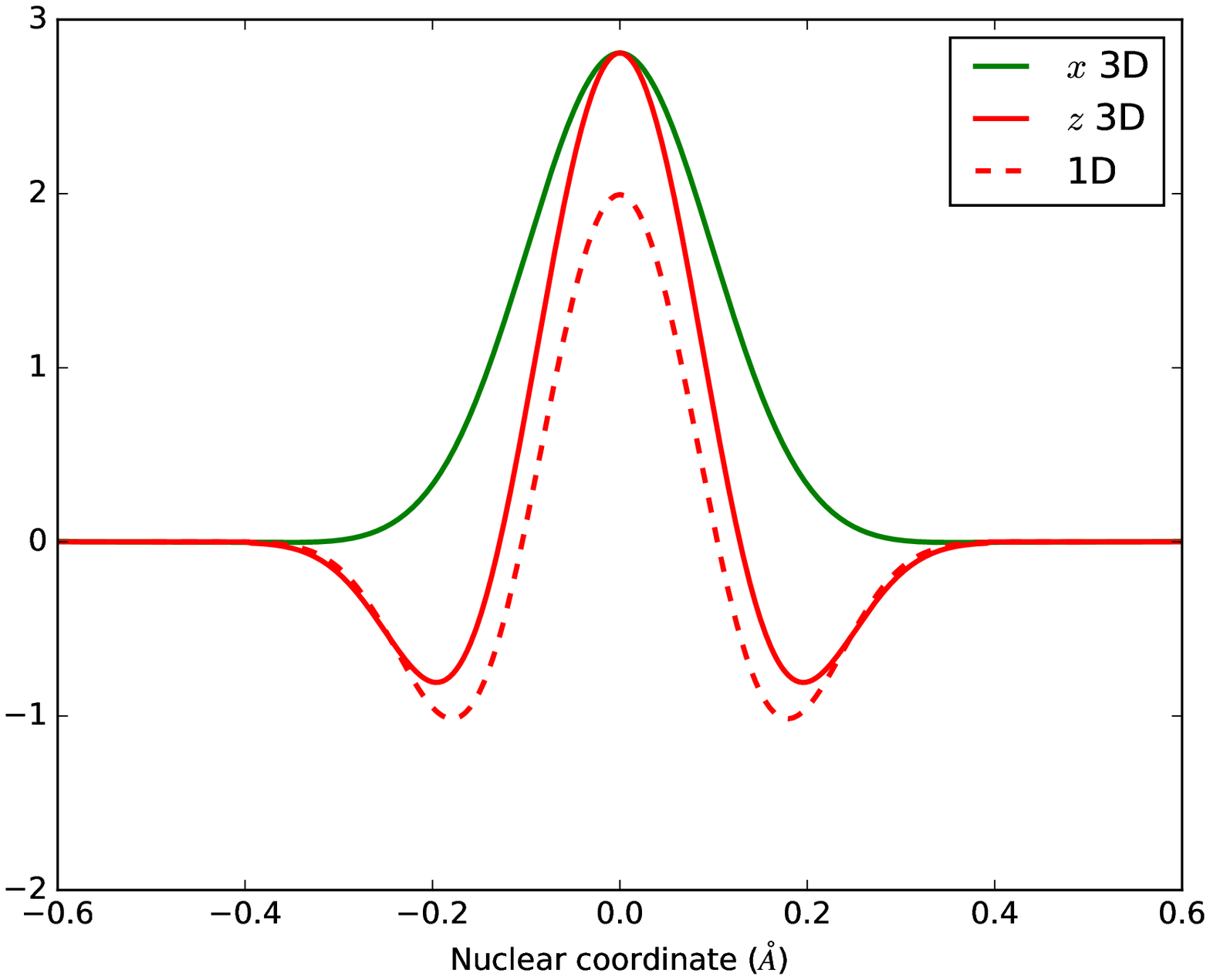}
}
\subfloat[\label{fig:FHF_3DNucDen1}]{
   \includegraphics[width=0.5\textwidth]{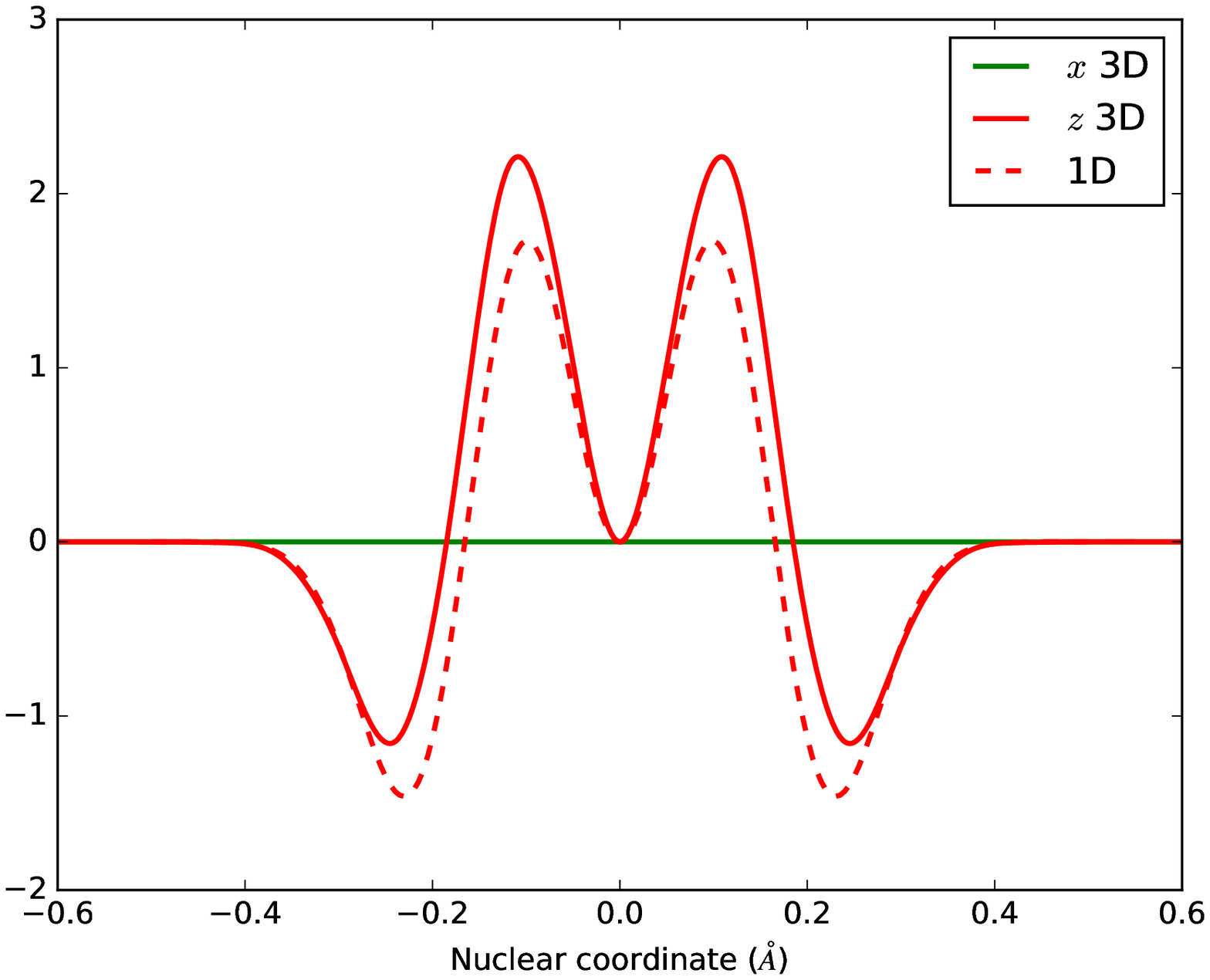}
}
\caption{The $x$ and $z$-component of 3D NECSCF corrections 
to RHF protonic density (mau) of FHF\textsuperscript{-} for $\nu$=0 \textbf{(a)} and $\nu$=1 \textbf{(b)} 
vibrational levels. The 1D corrections along $z$-direction is computed by freezing 
the $x$- and $y$-directions. The root-mean-square deviations in the proton density over 
3D space are 0.19 and 0.14 mau for $\nu=0$ and $\nu=1$, respectively.
The RHF protonic density is calculated using the BO RHF PES. The F-F distances
are determined according to the PES minima for NECSCF and RHF, respectively.
The $z$-component is along with the molecular axis whereas $x$ direction is
perpendicular to the molecular axis. 
}
        \label{fig:FHF_3DNucDen}
\end{figure*} 

{In Fig.~{\ref{fig:FHF_3DNucDen}} we present the slices of
three-dimensional (3D) NECSCF protonic difference density from the RHF
reference.  This quantity measures the response of protonic density from the
non-BO NECSCF correction for FHF$^-$ at $\nu=0$, for which the NECSCF energy
correction is computed to be 1.1 mau. 
Due to the rotational shape, the result for $y$-component is
the same as $x$. As opposed to Fig.~{\ref{fig:FHF_NucDen}}, the non-BO NECSCF
effect slightly enhances the protonic density around the equilibrium position
(-$0.15$~\AA~$-~0.15$~\AA) and depletes in the remaining region along the
molecular axis. At $\nu=1$, a nodal point appears at the equilibrium position,
and the protonic density enhancement area extends to -$0.2$~\AA~$-~0.2$~\AA. The
$x$-component does not show proton depletion at $n=0$, and it is constantly 0
since $\nu=1$ 3D proton density exhibits $xy$ nodal plane at the equilibrium
position. Interestingly, the one-dimensional (1D) nuclear difference density
computed by confining the protonic motion to $z$-direction demonstrates very
similar patterns with the $z$-component of 3D correction. For polyatomic
molecules with multiple vibrational degrees of freedom, the nucleus-electron
coupling behavior of individual mode can be examined by selecting the nuclear
motion that makes the most significant contribution to the non-BO correction.
}

\section{Conclusions}

The significance of non-BO effects involving nucleus-electron correlated motion
is manifested in many chemistry problems that are difficult for traditional
quantum chemistry methods within the BO framework.  We have developed an NECSCF
method for conveniently treating the nucleus-electron coupling using an exact
factorization of the molecular wavefunction.  The working equations for the
nuclear and electronic subsystems are derived and implemented currently for
uncorrelated electronic ground state for which the NECSCF theory is exact and
invariant to unitary orbital rotation, with the particular choice of  geometric
gauge transformation that yields an anti-symmetric derivative vector potential. The
post-HF electronic correlation can be computed by using resulting NECSCF MOs, as
demonstrated to MP2 and FCI correlation methods.  The computations of
vibrationally averaged properties, including molecular energy, bond length,
electronic and nuclear density are demonstrated and compared.  The NECSCF
predicts that nucleus-electron coupling generally weakens the bonding strength
and favors bond elongation, as well as an electron density transfer out of the
bonding area along the vibrational progression in which the nucleus-electron
coupling takes place. The present work not only provides an alternative
computational approach for accurately treating non-BO problems, but also lays
the basis for a general methodological framework in which systematic methods can
be developed for nucleus-coupled strongly correlated electrons and excited
electronic states on multiple PESs from NECSCF mean-field wavefunction in our
future endeavor.

\section*{Supplementary Material}

The supplementary material contains data for proton affinities and basis set
limit extrapolation results for H$_2^+$, HD$^+$, D$_2^+$ and H$_2$.

\begin{acknowledgments}
We acknowledge the funding support from the Seed Fund Program for Basic Research
(Grant No. 201711159116) by the University of Hong Kong and the Computational
Initiative provided by the Faculty of Science at HKU. J.Y. acknowledges the
research program of AIR@InnoHK cluster from the Innovation and Technology
Commission of Hong Kong SAR of China. The computations were partially performed
using research computing facilities offered by Information Technology Services,
the University of Hong Kong. We are grateful to Professor Sharon Hammes-Schiffer
for helpful discussions and comments.
\end{acknowledgments}

\section*{Data Availability Statement}

The data that supports the findings of this study are available within the
article and its supplementary material.

\appendix

{
\section{\label{sec:derivation}Derivation of form-invariant NECSCF electronic and nuclear equations}

The total molecular energy functional $\mathscr{L}$ in terms of
$\Phi(\mathbf{rs},\mathbf{R})$ and $\chi(\mathbf{R})$, subject to the
normalization conditions in the electronic and nuclear coordinate space, is
defined in Eq.~(\ref{eq:lagrang}). $\Phi(\mathbf{rs},\mathbf{R})$ here can be
any type of electronic wavefunctions, not limited to the mean-field determinant.
By variationally minimizing $\mathscr{L}$ with respect to
$\Phi(\mathbf{rs},\mathbf{R})$ and $\chi(\mathbf{R})$, the nonadiabatically
coupled electronic and nuclear equations have been shown form-invariant under
gauge transformation,\cite{Gross2014,requist2016exact} but are not
energy-invariant in general. In our approach, $\Phi(\mathbf{rs},\mathbf{R})$ is
assumed as a Slater determinant and $\mathscr{L}$ is minimized with respect to
non-BO MOs $\psi_i(\mathbf{rs},\mathbf{R})$ to obtain the NECSCF electronic
working equation. We adopt the spin orbital notation in the derivation and
$\mathbf{s}$ coordinate is dropped for simplicity.  
By evaluating
$\mel**{\Psi(\mathbf{r},\mathbf{R})}{\hat{H}}{\Psi(\mathbf{r},\mathbf{R})}_{\mathbf{r}}$,
the energy functional $\mathscr{L}$ in terms of non-BO spatial orbital
$\psi_i(\mathbf{r},\mathbf{R})$ and $\chi(\mathbf{R})$ reads,
\begin{widetext}
\begin{eqnarray}
\mathscr{L} &=&
\int\text{d}\mathbf{R}\chi^\dagger(\mathbf{R})\chi(\mathbf{R})
\mel**{\Phi(\mathbf{r},\mathbf{R})}{\hat{H}_\text{BO}}{\Phi(\mathbf{r},\mathbf{R})}_\mathbf{r}
-\int\text{d}\mathbf{R}\sum_\nu\frac{\chi^\dagger(\mathbf{R})\grad_\nu\chi(\mathbf{R})}{M_\nu}
\sum_i\mel**{\overline\psi_i}{\grad_\nu}{\psi_i}
\nonumber \\ 
&&-\int\text{d}\mathbf{R}\chi^\dagger(\mathbf{R})\chi(\mathbf{R})\sum_\nu\frac{1}{2M_\nu}
\Big[\sum_i\mel**{\overline\psi_i}{\grad^2_\nu}{\psi_i} +
\sum_i\sum_{j}\mel**{\overline\psi_i}{\grad_\nu}{\psi_i}
\cdot\mel**{\overline\psi_j}{\grad_\nu}{\psi_j} \nonumber \\
&& -\mel**{\overline\psi_i}{\grad_\nu}{\psi_j}\cdot\mel**{\overline\psi_j}{\grad_\nu}{\psi_i}\Big] 
+
\int\text{d}\mathbf{R}\chi^\dagger(\mathbf{R})\hat{T}_n\chi(\mathbf{R})\braket{\Phi(\mathbf{r},\mathbf{R})}{\Phi(\mathbf{r},\mathbf{R})}
\nonumber \\
&&-\int\text{d}\mathbf{R}\sum_i\sum_{j}\tilde
\epsilon_{ij}(\mathbf{R})\left(\braket{\overline\psi_i}{\psi_j} -
\delta_{ij}\right)
- E\left[\braket{\chi(\mathbf{R})}{\chi(\mathbf{R})}_\mathbf{R}-1\right].
\label{eq:lagrangmo}
\end{eqnarray}
\end{widetext}
The biorthogonality constrain $\braket{\overline{\psi}_i}{\psi_j} - \delta_{ij}
=0$ ensures the PNC, and replaces the electronic normalization condition
$\braket{\Phi(\mathbf{r},\mathbf{R})}{\Phi(\mathbf{r},\mathbf{R})}_{\mathbf{r}}-1=0$
in Eq.~(\ref{eq:lagrang}).  The Lagrangian energy of Eq.~(\ref{eq:lagrangmo}) is
variationally minimized with respect to $\bra{\overline{\psi}_i}$,
$\chi(\mathbf{R})$, the multipliers $\tilde\epsilon_{ij}(\mathbf{R})$ and $E$,
which leads to the NECSCF coupled electronic and nuclear equations as follows,
\begin{equation}
\Big[\hat{F}_\mathrm{BO} + \hat{V}^\mathrm{cp}_\chi \Big]\ket{\psi_i} 
=\frac{1}{|\chi(\mathbf{R})|^2}\sum_{j}\tilde\epsilon_{ij}(\mathbf{R})\ket{\psi_j},
\label{eq:working_elec}
\end{equation}
\begin{equation}
\Big[\hat{T}_n + \hat{E}_{el}(\mathbf{R}) \Big]\chi = E\chi.  
\label{eq:working_nuc}
\end{equation}
Above, 
the embedding operator $\hat{V}^\mathrm{cp}_\chi$ is formally exact for
uncorrelated electrons,
\begin{equation}
 \hat{V}^\mathrm{cp}_\chi = -\sum_\nu\frac{1}{M_\nu}\left(
\frac{\grad_\nu\chi}{\chi} \cdot\grad_\nu + \frac{\grad^2_\nu}{2}
+\sum_{j}\braket{\overline{\psi}_j}{\grad_\nu \psi_j}\cdot \grad_\nu
-\ket{\grad_\nu \psi_j} \cdot \bra{\overline{\psi}_j}\grad_\nu\right).
\end{equation}

In the nuclear equation, $\hat{E}_{el}(\mathbf{R})$ provides the NECSCF electronic
potential operator,
\begin{eqnarray}
\hat{E}_{el}(\mathbf{R})  &=& 
\mel{\Phi(\mathbf{r},\mathbf{R})}{\hat{H}_\mathrm{BO}}{\Phi(\mathbf{r},\mathbf{R})}_{\mathbf{r}}
-\sum_\nu\frac{1}{2M_\nu}\sum_i\Big(
\mel**{\overline\psi_i}{\grad^2_\nu}{\psi_i} 
+2\mel**{\overline\psi_i}{\grad_\nu}{\psi_i} \cdot\grad_\nu
\nonumber \\
&&+ \sum_{j}\mel**{\overline\psi_i}{\grad_\nu}{\psi_i}
\cdot\mel**{\overline\psi_j}{\grad_\nu}{\psi_j}
-\mel**{\overline\psi_i}{\grad_\nu}{\psi_j}
\cdot\mel**{\overline\psi_j}{\grad_\nu}{\psi_i}\Big).
\label{eq:ene1}
\end{eqnarray}
By solving the electronic equation of Eq.~(\ref{eq:working_elec}), we can
evaluate the NECSCF one-electron energy correction $\delta \epsilon$,
\begin{eqnarray}
 \delta \epsilon(\mathbf{R}) &=& \sum_i
\matrixel{\overline\psi_i}{\hat{V}_\chi^\mathrm{cp}}{\psi_i}.
\label{eq:deltaoneE}
\end{eqnarray}
The NECSCF energy correction $\delta E_{el}(\mathbf{R})$ is
\begin{eqnarray}
 \delta E_{el}(\mathbf{R}) &=& \sum_i
\matrixel{\overline\psi_i}{\left(\hat{V}_\chi^\mathrm{cp} 
 +\frac{1}{2}\sum_\nu\frac{1}{M_\nu}\sum_j\braket{\overline{\psi}_j}{\grad_\nu \psi_j}\cdot \grad_\nu
 -\ket{\grad_\nu \psi_j} \cdot \bra{\overline{\psi}_j}\grad_\nu\right)}{\psi_i}.
\label{eq:deltaE}
\end{eqnarray}
 
Thus, the geometrically averaged NECSCF electronic energy $E_{el}$
contains both contributions
from the averaged BO electronic energy $E_\mathrm{BO}$ and the NECSCF correction,
\begin{equation}
E_{el} = \matrixel{\chi}{\hat{E}_{el}(\mathbf{R})}{\chi}_\mathbf{R}= E_\mathrm{BO} +
\matrixel{\chi}{\delta E_{el}(\mathbf{R})}{\chi}_\mathbf{R}
\end{equation}

Next, we show that Eq.~(\ref{eq:working_elec}), the NECSCF electronic equation,
is form-invariant and independent of the gauge choice. 
To this end, we only need to show that
the last two terms of the embedding operator $\hat{V}_\chi^\mathrm{cp}$ 
are form-invariant, since only these terms are explicitly related to NECSCF MOs.
Using the spectral representation
$\hat{V}_\chi^\mathrm{cp}=\sum_{pq}\ket{\psi_p}V_{\chi,pq}^\mathrm{cp}\bra{\overline\psi_q}$
in the complete NECSCF MOs, the last two terms of
$\hat{V}_\chi^\mathrm{cp}$ in Eq.~(\ref{eq:vc}) read
\begin{eqnarray}
\sum_j\matrixel{\overline{\psi}_j}{\grad_\nu}{\psi_j}\cdot \grad_\nu   &=&
\sum_{pq}\ket{\psi_p} \sum_j\matrixel{\overline{\psi}_j}{\grad_\nu}{\psi_j}\cdot
\matrixel{\overline\psi_p}{\grad_\nu}{\psi_q}  \bra{\overline\psi_q} \nonumber 
\\
&=& 
\sum_{pq}\ket{\psi_p} \tr(\mathbf{A}^{(\nu)}_\mathrm{oo})\cdot
A_{pq}^{(\nu)}\bra{\overline\psi_q}, \\
\label{eq:form1}
\sum_j\ket{\grad_\nu \psi_j} \cdot \bra{\overline{\psi}_j}\grad_\nu
&=& \sum_j\sum_{pq}\ket{\psi_p} \matrixel{\overline\psi_p}{\grad_\nu}{\psi_j} \cdot
\matrixel{\overline\psi_j}{\grad_\nu}{\psi_q}\bra{\overline\psi_q} 
\nonumber \\
&=& \sum_{pq}\ket{\psi_p} \sum_j A_{pj}^{(\nu)} \cdot A_{jq}^{(\nu)} \bra{\overline\psi_q}.
\label{eq:form2}
\end{eqnarray}
According to the unitary invariance given in Eqs.~(\ref{eq:Apoo})
and~(\ref{eq:Aptr}), these operators are form-invariant for a unitary
transformation of NECSCF MOs. We can then pick NECSCF MOs that lead to a formal
orbital energy $\epsilon_{i}=\delta_{ij}\sum_{i' j'} T^*_{i'
i}\tilde\epsilon_{i' j'}T_{j' j}$ by a unitary transformation of the multipliers
$\tilde\epsilon_{ij}$. 

\section{\label{sec:effective}Effective coupled equations}

The nuclear equation~(\ref{eq:working_nuc}) indicates that the NECSCF electronic
potential operator $\hat{E}_{el}(\mathbf{R})$ in Eq.~(\ref{eq:ene1}) yields a
potential surface that depends on the nuclear wavefunction, unless the special
gauge choice making $A_{ii}^{(\nu)}=0$ is applied according to
Eqs.~(\ref{eq:gaug21}) and~(\ref{eq:gaug22}). To better understand the nature
of the coupled nuclear motion and associated energy surface, we define an
effective geometric derivative
$\widetilde\grad_\nu$ for the $\nu$-th vibration
\begin{equation}
 \widetilde\grad_\nu = \grad_\nu + \tr(\mathbf{A}_\mathrm{oo}^{(\nu)}),
\label{eq:grad2}
\end{equation}
where the geometric derivative matrix $\mathbf{A}_\mathrm{oo}^{(\nu)}$ is
composed of elements from the vector
$\matrixel{\overline\psi_i}{\grad_\nu}{\psi_j}$ for occupied MOs.
The electronic equation is then converted to an equivalent form,
\begin{equation}
\Big[\frac{}{}\hat{F}_\mathrm{BO} +\delta\epsilon
+\hat{\widetilde{V}}^\mathrm{cp}_\chi \Big]
\ket{\psi_i} = \frac{ \epsilon_i}{|\chi(\mathbf{R})|^2} \ket{\psi_i},
\label{eq:effect_elec} 
\end{equation}
with the effective embedding potential $\hat{\widetilde V}_\chi^\mathrm{cp}$ 
\begin{equation}
\hat{\widetilde V}_\chi^\mathrm{cp} =
-\sum_\nu\frac{1}{M_\nu}\Big(
\frac{\widetilde\grad_\nu\chi}{\chi} \cdot\widetilde\grad_\nu +
\frac{{\widetilde\grad}^2_\nu}{2}
-\sum_{j} \ket{\widetilde\grad_\nu \psi_j} \cdot
\bra{\overline{\psi}_j}\widetilde\grad_\nu
\Big),
\end{equation}
and an orbital energy shift $\delta\epsilon$,
\begin{equation}
\delta\epsilon = \sum_\nu \frac{1}{M_\nu}\left[
-\frac{1}{2}\tr^2(\mathbf{A}^{(\nu)}_\mathrm{oo})
%-\tr^{}(\mathbf{A}^{(\nu)}_\mathrm{oo})\cdot \widetilde\grad_\nu
+\tr^{}(\mathbf{A}_\mathrm{oo}^{(\nu)})\frac{\widetilde\grad_\nu\chi}{\chi}
\right].
\end{equation}
Similarly, the effective nuclear equation is given by
\begin{equation}
\Big[\hat{\widetilde T}_{n} +\widetilde E_{el}(\mathbf{R})\Big]\chi= E\chi
\label{eq:effect_nuc}
\end{equation}
where the effective kinetic energy operator $\hat{\widetilde T}_\nu$ is
\begin{equation}
\hat{\widetilde T}_n = -\sum_\nu \frac{\widetilde\grad_\nu^2}{2M_\nu}
\label{eq:tn2}
\end{equation}
and the effective electronic energy $\widetilde E_{el}(\mathbf{R})$
\begin{eqnarray}
\widetilde E_{el}(\mathbf{R})&=&
\mel{\Phi(\mathbf{r},\mathbf{R})}{\hat{H}_\mathrm{BO}}{\Phi(\mathbf{r},\mathbf{R})}_{\mathbf{r}}
+\sum_i\matrixel{\overline\psi_i}{\left(\delta\epsilon + \hat{\widetilde
V}_\chi^\mathrm{cp}{-\frac{1}{2}\sum_\nu\frac{1}{M_\nu}\sum_{j} \ket{\grad_\nu \psi_j} \cdot
\bra{\overline{\psi}_j}\grad_\nu}\right)}{\psi_i}
\nonumber \\
&=&
\mel{\Phi(\mathbf{r},\mathbf{R})}{\hat{H}_\mathrm{BO}}{\Phi(\mathbf{r},\mathbf{R})}_{\mathbf{r}}
-\sum_\nu\frac{1}{2M_\nu}\Big(\sum_i\mel**{\overline\psi_i}{\widetilde\grad^2_\nu}{\psi_i}
-\sum_{ij}\mel**{\overline\psi_i}{\widetilde\grad_\nu}{\psi_j}
\cdot\mel**{\overline\psi_j}{\widetilde\grad_\nu}{\psi_i}\Big)
\nonumber \\
\label{eq:ene2}
\end{eqnarray}
which is analogous to Eq.~(\ref{eq:deltaE}).
In deriving Eq.~(\ref{eq:ene2}), we note that the first term of
$\delta\epsilon$ make no additive contribution to $\widetilde
E_{el}(\mathbf{R})$, and the third term compensates
$-\frac{\widetilde\grad_\nu\chi}{\chi}\cdot \widetilde\grad_\nu$ from
$\hat{\widetilde V}_\chi^\mathrm{cp}$. It becomes clear that the nuclear motion
carrying $\hat{\widetilde{T}_n}$ can be described on an effective electronic
energy surface that is computed as the the NECSCF electronic energy correction.
}

\section{\label{sec:unitary}Energy invariance to unitary rotation of NECSCF MOs}

From the NECSCF energy correction $\delta E_{el}(\mathbf{R})$ in Eq.~(\ref{eq:deltaE}),
the NECSCF electronic energy contribution can be further formulated in tracing matrix,
\begin{eqnarray}
 \delta E_{el}(\mathbf{R}) &=&
 -\sum_\nu\frac{1}{M_\nu}
\left(
\frac{\grad_\nu\chi}{\chi} \cdot A_{ii}^{(\nu)}+ \frac{1}{2}\sum_a A_{ia}^{(\nu)}\cdot
A_{ai}^{(\nu)} + \frac{1}{2}\grad_\nu \cdot A_{ii}^{(\nu)}
\right)
\nonumber \\
&=&-\sum_\nu \frac{1}{2M_\nu}\left(\frac{\grad_\nu\chi}{\chi}
\tr(\mathbf{A}_\mathrm{oo}^{(\nu)})+\tr(\mathbf{A}_\mathrm{ov}^{(\nu)}\mathbf{A}_\mathrm{vo}^{(\nu)})\right)
\label{eq:delE}
\end{eqnarray}
We will show that all these terms lead to $\delta E_{el}$ that is energy 
invariant to an unitary orbital rotation among all occupied MOs, regardless of
the gauge choice.
The matrices $\mathbf{A}_\mathrm{oo}^{(\nu)}$, $\mathbf{A}_\mathrm{ov}^{(\nu)}$ and
$\mathbf{A}_\mathrm{vo}^{(\nu)}$ collect the elements $A_{ij}^{(\nu)}$, $A_{ia}^{(\nu)}$ and
$A_{ai}^{(\nu)}$, and are generically evaluated as follows, respectively,
\begin{eqnarray}
 \mathbf{A}_\mathrm{oo}^{(\nu)} &=&
\mathbf{\overline C}^\dagger_\mathrm{o}\mathbf{S}^{0\nu}\mathbf{C}_\mathrm{o}+\mathbf{U}_\mathrm{oo}^{(\nu)}
\label{eq:Aoo}
\\
 \mathbf{A}_\mathrm{ov}^{(\nu)} &=&
\mathbf{\overline C}^\dagger_\mathrm{o}\mathbf{S}^{0\nu}\mathbf{C}_\mathrm{v}+\mathbf{U}_\mathrm{ov}^{(\nu)}
\\
 \mathbf{A}_\mathrm{vo}^{(\nu)} &=&
\mathbf{\overline C}^\dagger_\mathrm{v}\mathbf{S}^{0\nu}\mathbf{C}_\mathrm{o}+\mathbf{U}_\mathrm{vo}^{(\nu)}
\end{eqnarray}
Consider the biorthogonal unitary transformations $\mathbf{T}_\mathrm{oo}$ and
$\mathbf{T}_\mathrm{vv}$ among the occupied and virtual MOs at any
molecular geometry $\mathbf{R}$, such that
$\mathbf{T}_\mathrm{oo}\mathbf{\overline T}_\mathrm{oo}^\dagger=\mathbf{I}$ and
$\mathbf{T}_\mathrm{vv}\mathbf{\overline T}_\mathrm{vv}^\dagger=\mathbf{I}$,
respectively.  The transformed MOs are
\begin{eqnarray}
 \mathbf{C}^\prime_\mathrm{o} &=&
 \mathbf{C}_\mathrm{o}\mathbf{T}_\mathrm{oo},
\label{eq:t1}\\
 \mathbf{C}^\prime_\mathrm{v} &=&
 \mathbf{C}_\mathrm{v}\mathbf{T}_\mathrm{vv},
\label{eq:t2}
\end{eqnarray}
The geometrically perturbed (e.g., $\mathbf{C}^\prime$ and $\mathbf{C}$) and the unperturbed
(e.g., $\mathbf{C}^{\prime(0)}$ and $\mathbf{C}^{(0)}$) MOs are related by,
\begin{eqnarray}
 \mathbf{C}^\prime_\mathrm{o}&=&
  \mathbf{C}^{\prime(0)}_\mathrm{o}\mathbf{U}_\mathrm{oo}^{\prime}
+ \mathbf{C}^{\prime(0)}_\mathrm{v}\mathbf{U}_\mathrm{vo}^{\prime},
\label{eq:o1}\\
 \mathbf{C}^\prime_\mathrm{v} &=&
 \mathbf{C}^{\prime(0)}_\mathrm{v}\mathbf{U}_\mathrm{vv}^{\prime}
+\mathbf{C}^{\prime(0)}_\mathrm{o}\mathbf{U}_\mathrm{ov}^{\prime}.
\label{eq:v1}\\
 \mathbf{C}_\mathrm{o} &=&
 \mathbf{C}^{(0)}_\mathrm{o}\mathbf{U}_\mathrm{oo}
+\mathbf{C}^{(0)}_\mathrm{v}\mathbf{U}_\mathrm{vo},
\label{eq:o2}\\
 \mathbf{C}_\mathrm{v}&=&
 \mathbf{C}^{(0)}_\mathrm{v}\mathbf{U}_\mathrm{vv}
+\mathbf{C}^{(0)}_\mathrm{o}\mathbf{U}_\mathrm{ov}.
\label{eq:v2}
\end{eqnarray}
Combining Eqs.~(\ref{eq:t1})--(\ref{eq:v2}) and noting that
$\mathbf{U}^{(0)}_\mathrm{oo}=\mathbf{U}^{(0)}_\mathrm{vv}=\mathbf{I}$
and
$\mathbf{U}^{(0)}_\mathrm{ov}=\mathbf{U}^{(0)}_\mathrm{vo}=0$,
we arrive at the following relations between the transformed and original
relaxations,
\begin{eqnarray}
 \mathbf{U}^{\prime(\nu)}_\mathrm{oo}&=&
\mathbf{T}^{\dagger(0)}_\mathrm{oo}\mathbf{U}^{(\nu)}_\mathrm{oo}\mathbf{T}^{(0)}_\mathrm{oo}
+\left(\mathbf{T}^{\dagger(0)}_\mathrm{oo}\mathbf{T}^{(0)}_\mathrm{oo}\right)^{(\nu)}
= \mathbf{T}^{\dagger(0)}_\mathrm{oo}\mathbf{U}^{(\nu)}_\mathrm{oo}\mathbf{T}^{(0)}_\mathrm{oo},
\label{eq:uoo}\\
 \mathbf{U}^{\prime(\nu)}_\mathrm{ov}&=&
\mathbf{T}^{\dagger(0)}_\mathrm{oo}\mathbf{U}^{(\nu)}_\mathrm{ov}\mathbf{T}^{(0)}_\mathrm{vv},
\label{eq:uov}\\
 \mathbf{U}^{\prime(\nu)}_\mathrm{vo}&=&
\mathbf{T}^{\dagger(0)}_\mathrm{vv}\mathbf{U}^{(\nu)}_\mathrm{vo}\mathbf{T}^{(0)}_\mathrm{oo}.
\label{eq:uvo}
\end{eqnarray}

Now using the relations of Eqs.~(\ref{eq:Aoo}) and~\ref{eq:uoo}, 
we can evaluate $\mathbf{A}^{\prime(\nu)}_\mathrm{oo}$ and
$\tr(\mathbf{A}^{\prime(\nu)}_\mathrm{oo})$,
\begin{eqnarray}
\mathbf{A}^{\prime(\nu)}_\mathrm{oo} & = & 
\mathbf{\overline{T}}^{\dagger(0)}_\mathrm{oo}\mathbf{A}^{(\nu)}_\mathrm{oo}\mathbf{T}^{(0)}_\mathrm{oo}
\label{eq:Apoo}
\\
\tr(\mathbf{A}^{\prime(\nu)}_\mathrm{oo}) &=&
\tr(\mathbf{\overline{T}}^{\dagger(0)}_\mathrm{oo}\mathbf{A}^{(\nu)}_\mathrm{oo}\mathbf{T}^{(0)}_\mathrm{oo})
%\tr(\mathbf{S}^{0\nu}\mathbf{C}^\prime_\mathrm{o}\mathbf{\overline C}^{\prime\dagger}_\mathrm{o})
%\tr(\mathbf{U}_\mathrm{oo}^{\prime(\nu)}) 
\nonumber \\
&=&
%\tr(\mathbf{S}^{0\nu}\mathbf{C}_\mathrm{o}\mathbf{\overline C}^\dagger_\mathrm{o})
\tr(\mathbf{A}^{(\nu)}_\mathrm{oo}\mathbf{T}^{(0)}_\mathrm{oo}\mathbf{T}^{\dagger(0)}_\mathrm{oo})
%+\tr(\mathbf{T}^{\dagger(0)}_\mathrm{oo}\mathbf{T}^{(\nu)}_\mathrm{oo})
\nonumber \\
&=&
\tr{\mathbf{A}^{(\nu)}_\mathrm{oo}}.
%+\tr{\mathbf{T}^{\dagger(0)}_\mathrm{oo}\mathbf{T}^{(\nu)}_\mathrm{oo}}.
\label{eq:Aptr}
\end{eqnarray}
It is obvious that $\tr(\mathbf{A}^{\prime(\nu)}_\mathrm{oo})$ is invariant
regardless of the gauge choice.
%unless this term is zero by the gauge choice or excluded in the NECSCF energy evaluation.

For $\tr(\mathbf{A}_\mathrm{ov}^{(\nu)}\mathbf{A}_\mathrm{vo}^{(\nu)})$, there
is
\begin{eqnarray}
\tr(\mathbf{A}_\mathrm{ov}^{(\nu)}\mathbf{A}_\mathrm{vo}^{(\nu)})&=&
\tr(
\mathbf{S}^{0\nu}\mathbf{C}_\mathrm{v}
\mathbf{\overline C}^\dagger_\mathrm{v}\mathbf{S}^{0\nu}\mathbf{C}_\mathrm{o}
\mathbf{\overline C}^\dagger_\mathrm{o}
+\mathbf{U}_\mathrm{ov}^{(\nu)}\mathbf{U}_\mathrm{vo}^{(\nu)}
+ \mathbf{S}^{0\nu}\mathbf{C}_\mathrm{v}
 \mathbf{U}_\mathrm{vo}^{(\nu)}\mathbf{\overline C}^\dagger_\mathrm{o}
+ \mathbf{C}_\mathrm{o}\mathbf{U}_\mathrm{ov}^{(\nu)}
\mathbf{\overline C}^\dagger_\mathrm{v}\mathbf{S}^{0\nu}
).
\end{eqnarray}
By referring to the relations in
Eqs.~(\ref{eq:t1}),~(\ref{eq:t2}),~(\ref{eq:uov}) and~(\ref{eq:uvo}), all the
four terms are intrinsically invariant due to the unitarity of
transformation matrices, which leads to
\begin{eqnarray}
\tr(\mathbf{A}_\mathrm{ov}^{\prime(\nu)}\mathbf{A}_\mathrm{vo}^{\prime(\nu)}) &=&
\tr(\mathbf{A}_\mathrm{ov}^{(\nu)}\mathbf{A}_\mathrm{vo}^{(\nu)})
\end{eqnarray}
for any gauge transformation.

The invariance of the electronic potential leaves the formulation of
nuclear equation in Eq.~(\ref{eq:working2}) 
invariant, i.e., both the total molecular energy and the nuclear density
amplitudes on the same uniform grid set do not change.
Therefore the occupied unitary $\mathbf{T}_\mathrm{oo}$ transforms
the determinant wavefunction and the total wavefunction $\Psi$ by a phase factor
$\det(\mathbf{T}_\mathrm{oo})=e^{i\theta(\mathbf{R})}$,
\begin{eqnarray}
\Phi^\prime &=& \Phi \det(\mathbf{T}_\mathrm{oo}), \\
\Psi^\prime &=& \chi\Phi^\prime = \text{det}(\mathbf{T}_\mathrm{oo})\Psi = e^{i\theta(\mathbf{R})}\Psi.
\end{eqnarray}

\section{\label{sec:fghgrad}Analytical first- and second-derivatives of FGH
nuclear wavefunction}

Assuming independent vibrational modes, we consider $\grad_\nu$ operator acting
on a FGH vibrational wavefunction $\chi_\nu$, leaving other vibrations frozen.
The Fourier representation of the Dirac delta function is
\begin{eqnarray}
\delta(\mathbf{R}_\nu^\prime-\mathbf{R}_\nu) = 
\frac{1}{2\pi}\int_{-\infty}^{+\infty} e^{-ik(\mathbf{R}_\nu^\prime-\mathbf{R}_\nu)}\text{d}k
\end{eqnarray}
where the ranges of momentum space and position space are set to
$-\infty \leqslant k \leqslant +\infty$ and $0 \leqslant \mathbf{R}_\nu \leqslant +\infty$, respectively. Therefore,
\begin{eqnarray} \label{eq:nucgradfirst1}
\grad_\nu\chi_\nu &=& \frac{\partial}{\partial \mathbf{R}_\nu}
\int_{0}^{+\infty}\delta(\mathbf{R}_\nu^\prime-\mathbf{R}_\nu)\chi_\nu^\prime\text{d}R^\prime_\nu
\nonumber \\
&=&  \frac{\partial}{\partial \mathbf{R}_\nu}\int_{0}^{+\infty}
\left[\frac{1}{2\pi}\int_{-\infty}^{+\infty} 
e^{-ik(\mathbf{R}_\nu^\prime-\mathbf{R}_\nu)}\text{d}k\right]\chi_\nu^\prime\text{d}\mathbf{R}^\prime_\nu 
\nonumber \\
&=&  \int_{0}^{+\infty}\left[\frac{1}{2\pi}\int_{-\infty}^{+\infty} ik 
e^{-ik(\mathbf{R}_\nu^\prime-\mathbf{R}_\nu)}\text{d}k\right]\chi_\nu^\prime\text{d}\mathbf{R}^\prime_\nu.
\end{eqnarray}
Both $k$ and $\mathbf{R}_\nu$ in Eq.~(\ref{eq:nucgradfirst1}) are discretized on
uniform grid points. Suppose 
the length of the coordinate grid is $L$ and there are $N$ grid points ($N$ is odd), the spacings in
the position and momentum grid are given by $\Delta \mathbf{R}_\nu = \frac{L}{N}$ and
$\Delta k = \frac{2\pi}{L}$.\cite{FGH1989} The values of $k$ are evenly
distributed around the origin of zero, ranging from $-n\Delta k$ to $n\Delta k$
with $n = \frac{N-1}{2}$. The details of the discretization process are
available in ref.~\citenum{FGH2003}. Briefly, the matrix element of the
first-order derivative operator over the internal coordinates $(r_1,r_2)$
associated with the vibration $\chi_\nu$ defined on this grid set can be written
as 
\begin{eqnarray} \label{eq:nucgradfirst2}
\left[\grad_\nu\right]_{r_1r_2} = -\frac{4\pi}{LN}\sum_{k=1}^{n}k\text{sin}\left[\frac{2\pi k(r_1-r_2)}{N}\right].
\end{eqnarray}
Obviously, the matrix for the first-derivative operator is anti-Hermitian.  The
sum in Eq.~(\ref{eq:nucgradfirst2}) could be calculated analytically,
\begin{widetext}
\begin{eqnarray} \label{eq:nucgradfirst3}
\left[\grad_\nu\right]_{r_1r_2} = \begin{cases}
0 &\text{for $r_1 = r_2$,}\\
-\frac{2\pi}{LN}\left[\frac{\text{sin}[(n+1)(r_1-r_2)\frac{2\pi}{N}]}{2\text{sin}^2[(r_1-r_2)\frac{\pi}{N}]} 
+ \frac{(-1)^{r_1-r_2+1}(n+1)}{\text{sin}[(r_1-r_2)\frac{\pi}{N}]}\right] &\text{for $r_1 \ne r_2$.}
\end{cases}
\end{eqnarray}
\end{widetext}
A similar derivation leads to the grid representation of the second-derivative operator,
\begin{widetext}
\begin{eqnarray}\label{eq:nucgradsecond}
\left[\grad^2_\nu\right]_{r_1r_2} = 
\begin{cases}
-\frac{4\pi^2}{L^2}\frac{n(n+1)}{3} &\text{for $r_1 = r_2$,} \\
-\frac{4\pi^2}{L^2}\frac{(-1)^{r_2-r_1}(n+1)\text{cos}[(r_2-r_1)\frac{\pi}{N}]  
+ (n+1)\text{cos}[(n+1)(r_2-r_1)\frac{2\pi}{N}]
-\text{sin}[(n+1)(r_2-r_1)\frac{2\pi}{N}]\text{cot}[(r_2-r_1)\frac{\pi}{N}]}{2N\text{sin}^2[(r_2-r_1)\frac{\pi}{N}]} 
&\text{for $r_1 \ne r_2$.} 
\end{cases} 
\end{eqnarray}
\end{widetext}
The analytical first- (or second-) derivative of FGH nuclear wavefunction is
hence the matrix product between Eq.~(\ref{eq:nucgradfirst3}) (or
Eq.~(\ref{eq:nucgradsecond})) with the grid representation of the nuclear
wavefunction.

\nocite{*}
\bibliography{manuscript}% Produces the bibliography via BibTeX.

\end{document}